\documentclass[prb,twocolumn,superbib,tightenlines,narrowtext,showpacs]{revtex4}
\usepackage{amsfonts,amsmath,amssymb,amsthm,bm}
\usepackage{graphicx}
\begin{document}
\bibliographystyle{apsrev}
\title{Effect of Thermoelectric Cooling in Nanoscale Junctions}
\author{Yu-Shen Liu }
\author{Bailey C. Hsu}
\author{Yu-Chang Chen}
\email{yuchangchen@mail.nctu.edu.tw}
\affiliation{Department of Electrophysics, National Chiao Tung University, 1001 Ta Hsueh
Road, Hsinchu, Taiwan 30010 }
\begin{abstract}

We propose a thermoelectric cooling device based on an atomic-sized
junction. Using first-principles approaches, we investigate the working conditions and the
coefficient of performance (COP) of an atomic-scale electronic refrigerator where
the effects of phonon's thermal current and local heating are included.
It is observed that the functioning of the thermoelectric nano-refrigerator is
restricted to a narrow range of driving voltages. Compared with the bulk
thermoelectric system with the overwhelmingly irreversible Joule heating, the
4-Al atomic refrigerator has a higher efficiency than a bulk thermoelectric
refrigerator with the same $ZT$ due to suppressed local heating via the
quasi-ballistic electron transport and small driving voltages. Quantum nature
due to the size minimization offered by atomic-level control of properties
facilitates electron cooling beyond the expectation of the conventional thermoelectric
device theory.

\end{abstract}
\pacs{73.50.Lw, 68.43.Pq, 73.40.Jn 81.07.Nb}
\maketitle
\section{Introduction}
The miniaturization of devices has been eliciting a tremendous wave of
multidisciplinary scientific interest~\cite{Aviram,book}. This interest is
motivated by the aspiration to develop new forms of electronic devices based
on nano-structures. To develop the nano-devices at the atomic/molecular level,
understanding of non-equilibrium quantum transport theory is of critical
importance. In the past decade, a growing number of studies have been
conducted to diversify the scopes of molecular electronics including the
current-voltage
characteristics~\cite{Kaun,DiVentra2,Nitzan1,Venkataraman,Gemma}, inelastic
electron tunneling spectroscopy
(IETS)~\cite{Wang2,Ratner3,Yluo,Kushmerick,Brandbyge,Dicarlo,cheniets,chenh2,thygesen}
, shot noise~\cite{Ruitenbeek1,Kiguchi,Natelson,chenshot}, counting
statistics~\cite{liushot}, local heating~\cite{chenheating,Huang2}, and
gate-controlled effect~\cite{DiVentragate,Ma,Solomon1,Solomon2,Reed}.
Substantial progress has been achieved in experiments and theories~\cite{Ahn,Lindsay,Tao1}.

Recently, atomic/molecular thermoelectric junctions are gaining increased
attention due to the recent measurements of the Seebeck coefficient, defined as
$S=dV/dT$, $dV$ is the voltage difference caused by the temperature difference
$dT$ by the Seebeck effect~\cite{Ruitenbeek,Majumdar1,Majumdar2,Majumdar3}.
Measurements of Seebeck coefficient provide a useful experimental approach to exploring the
electronic structure of the molecule bridging the electrodes~\cite{Malen}.
Methodologically, the scope of the research needs to extend through the
utilization of unprecedented experiments. These experiments inspire rapid
development in the theory of thermoelectricity at the atomic and molecular
scale~including the Seebeck coefficients, thermoelectric figure of merit
($ZT$), thermospin effect, and effect of electron-phonon
interactions~\cite{Paulsson,Zheng,Wang,Pauly,Galperin,Dubi,Markussen,Ke,Finch,Troels,Segal,Bergfield,Liu,Liu2,Hsu}%
.

Effect of thermoelectricity hybridizes the interactions between electron and
energy transport under non-equilibrium conditions. In the bulk and mesoscopic
systems, the efficiency of a thermoelectric (TE) refrigerator is usually
suppressed by a large work function. For example, the operation of a
thermoelectric cooling device such as a vacuum diode is limited to very high
temperatures ($T_{op}>1000$ K) due to its large potential
barrier~\cite{Mahan1}. Another pronounced drawback in the bulk system is
the overwhelmingly irreversible Joule heating due to diffused electrons which
significantly suppresses the efficiency of thermoelectric refrigerators.
Recently, researchers have looked into creating thermoelectric refrigerators that
operate at room temperatures. For this purpose, semiconductor
hetero-structures have been proposed to reduce the work
function~\cite{Mahan2,Chao,Dwyer,Westover}. A new solution to the low
temperature-operated thermoelectric refrigerators may be the atomic-sized
junctions, the extreme limit of device miniaturization.

The atomic-sized energy-conversion devices have gained growing interest in
material science and nanoscience. The electron transport mechanism in theses
systems is characterized by quasi-ballistic electron transport due to the small
size. Previous reports thus far have mainly focused on the Seebeck coefficient
($S$) and the thermoelectric figure of merit ($ZT$) of nanojunctions.
Several attempts have been made to understand the cooling mechanism in
nanojunctions~\cite{DiventraS,Galperin2,Segalreg1,Segalreg2}. This study
proposes a thermoelectric cooling device based on an atomic junctions. This
project develops a theory of the atomic-scale cooling mechanism for
quasi-ballistic electrons under non-equilibrium conditions from
first-principles approaches. We investigate the nano-refrigerator's working
conditions, the electron's thermal current which removes heat from the cold
temperature reservoir, and the coefficient of performance (COP) in the
presence of the phonon's thermal current and local heating. We observe that
the potential barrier is effectively suppressed by the resonant tunneling and
the local heating is significantly suppressed by the reduced dimension.
Nano-refrigerators with a figure of merit comparable to conventional TE
refrigerators usually have better coefficient of performance, taking advantage
the reduced local heating due to the small size. These quantum features
remarkably facilitates the electron cooling in nanoscale thermoelectric
refrigerators beyond the expectation of the conventional bulk thermoelectric
device theory.

\begin{figure}[ptb]
\includegraphics[width=7.5cm]{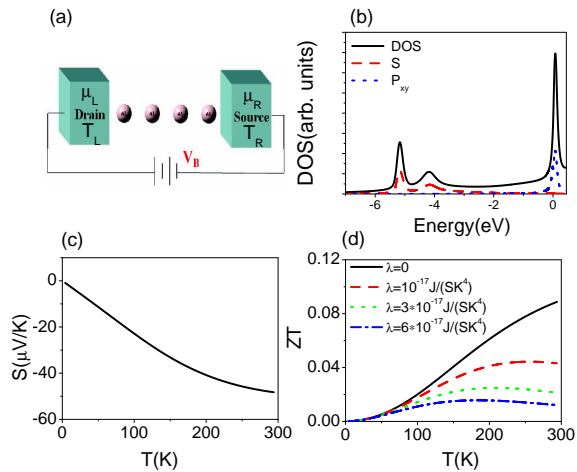}
\caption{
(color online)
(a) Scheme of the 4-Al atomic junction with
different temperatures $T_{L(R)}$ and chemical potentials $\mu_{L(R)}$. The
Al-jellium distance is about $2.5$~a.u. and the Al-Al bond distance is about
$6.3$~a.u.; (b) Density of state and projected density of states of the 4-Al
junction; (c) Seebeck coefficient vs. temperature of the 4-Al junction; and
(d) $ZT$ vs. T with the phonon's thermal current for various values of
$\lambda$. }%
\label{Fig1}
\end{figure}

As an example, we consider an ideal 4-Al monatomic junction, as depicted in
Fig.~\ref{Fig1}(a). The 4-Al atomic junction is
marked by a larger Seebeck coefficient induced by a sigma channel near the
chemical potentials with the $P_{x}-P_{y}$ orbital characters. Due to the small size,
local heating and photon radiation are sufficiently suppressed that they can
be safely neglected compared with the large phonon's thermal current
considered in this study. Calculations indicate that the 4-Al junction is able
to work at temperatures below 100 K. Compared with atomic-scale thermoelectric
power generators~\cite{liupowergen,Justin}, thermoelectric nano-refrigerators
require more stringent working conditions. For example, we observe that the
battery which drives thermoelectric nano-refrigerator is restricted to a range
of voltages between the lower and upper threshold biases.
The flow of the discussion in this paper is as follows. Firstly, in Sec.~II. we describe
the details of density functional theory, theory of thermoelectricity, and
theory of thermoelectric nano-refrigerator without and with the phonon's
thermal current and local heating.
In Sec.~III., we discuss the thermoelectric
properties of 4-Al atomic junctions, and subsequently discuss the onset bias,
working conditions, and COP of thermoelectric nano-refrigerators. We then
summarize our findings in Sec.~IV.
\section{Theoretical Methods}
The theory presented in the following is general to any
atomic/molecular junction characterized by the quasi-ballistic transport.
In Sec.~II.~ A., we present an
introduction of the density-functional theory (DFT). In
Sec.~II.~B., we present the theory to calculate
the Seebeck coefficient, electric conductance, electron's thermal conductance,
phonon's thermal conductance, and thermoelectric figure of merit ($ZT$). In
Sec.~II.~C., we present
the theory for electronic cooling without and with the phonon's thermal
current and local heating.
\subsection{Density Functional Theory}
We present a brief introduction of how to calculate the
electric current and the electron's thermal current in the DFT framework. We
picture a nanoscale junction as formed by two semi-infinite electrodes held
apart at a fixed distance with a nano-structured object bridging the gap
between them. The full Hamiltonian of the system is $H=H_{0}+V$, wherein
$H_{0}$ is the Hamiltonian due to the bare electrodes and $V$ is the
scattering potential of the nano-structured object.

First, we calculate the wave functions of the bare electrodes with an applied
bias $V_{B}=(\mu_{R}-\mu_{L})/e$, where $\mu_{L(R)}$ is the chemical potential
deep in the left (right) electrode. The unperturbed wave functions of the bare
electrodes have the form, $\Psi_{E\mathbf{K}_{\parallel}}^{0,L(R)}%
(\mathbf{r})=e^{i\mathbf{K}_{\parallel}\mathbf{\cdot R}}\cdot u_{E\mathbf{K}%
_{\parallel}}^{L(R)}(z)$, where $u_{E\mathbf{K}_{\parallel}}^{L(R)}(z)$
describes the electrons incident from the left (right) electrode before the
inclusion of the nano-structured object. The wavefunction $u_{E\mathbf{K}%
_{\parallel}}^{L(R)}(z)$ is calculated by solving the Shr\"{o}dinger equation
and Poisson equations iteratively until self-consistency is obtained. Note
that $u_{E\mathbf{K}_{\parallel}}^{R}(z)$ satisfies the following boundary
condition:
\begin{equation}
u_{E\mathbf{K}_{\parallel}}^{R}(z)=(2\pi)^{-\frac{3}{2}}\times%
\genfrac{\{}{.}{0pt}{}{\frac{1}{\sqrt{k_{R}}}(e^{-ik_{R}z}+R\text{ }%
e^{ik_{R}z}),\text{ }z\rightarrow\infty,}{\frac{1}{\sqrt{k_{L}}}T\text{
}e^{-ik_{L}z},\text{ }z\rightarrow-\infty,}
\label{uk2}%
\end{equation}
where $\mathbf{K}_{\parallel}$ is the electron momentum in the plane parallel
to the electrode surfaces, and $z$ is the coordinate parallel to the direction
of the current. The condition of energy conservation gives $\frac{1}{2}%
k_{R}^{2}=E-\frac{1}{2}\left\vert \mathbf{K}_{\parallel}\right\vert
^{2}-v_{eff}(\infty)$ and $\frac{1}{2}k_{L}^{2}=E-\frac{1}{2}\left\vert
\mathbf{K}_{\parallel}\right\vert ^{2}-v_{eff}(-\infty)$, where $v_{eff}(z)$
is the effective potential comprising the electrostatic and
exchange-correlation potentials.

The nano-structured object is considered as a scattering center. The
scattering wave functions of the entire system are calculated by solving the
Lippmann-Schwinger equation in the scattering approaches
iteratively until self-consistency is obtained,%
\begin{widetext}
\begin{equation}
\Psi_{E\mathbf{K}_{\parallel}}^{L(R)}(\mathbf{r})=\Psi_{E\mathbf{K}%
_{\parallel}}^{0,L(R)}(\mathbf{r})+\int d^{3}\mathbf{r}_{1}\int d^{3}%
\mathbf{r}_{2}G_{E}^{0}(\mathbf{r},\mathbf{r}_{1})V(\mathbf{r}_{1}%
,\mathbf{r}_{2})\Psi_{E\mathbf{K}_{\parallel}}^{L(R)}(\mathbf{r}_{2}),
\label{Lippmann2}%
\end{equation}
\end{widetext}
where $\Psi_{E\mathbf{K}_{\parallel}}^{L(R)}(\mathbf{r})$ stands for the
effective single-particle wave functions of the entire system, which also
represents the electrons with energy $E$ incident from the left (right)
electrode. The potential $V(\mathbf{r}_{1},\mathbf{r}_{2})$ that the electrons
experience when they scatter through the nanojunction is%
\begin{widetext}
\begin{equation}
V(\mathbf{r}_{1},\mathbf{r}_{2})=V_{ps}(\mathbf{r}_{1},\mathbf{r}%
_{2})+\left\{  \left(  V_{xc}\left[  n\left(  \mathbf{r}_{1}\right)  \right]
-V_{xc}\left[  n_{0}\left(  \mathbf{r}_{1}\right)  \right]  \right)  +\int
d\mathbf{r}_{3}\frac{\delta n\left(  \mathbf{r}_{3}\right)  }{\left\vert
\mathbf{r}_{1}-\mathbf{r}_{3}\right\vert }\right\}  \delta(\mathbf{r}%
_{1}-\mathbf{r}_{2}), \label{V2}%
\end{equation}
\end{widetext}
where $V_{ps}(\mathbf{r}_{1},\mathbf{r}_{2})$ is the electron-ion interaction
potential represented with pseudopotential; $V_{xc}\left[  n\left(
\mathbf{r}\right)  \right]  $ is the exchange-correlation potential calculated
at the level of the local-density approximation; $n_{0}\left(  \mathbf{r}%
\right)  $ is the electron density for the pair of biased bare electrodes;
$n\left(  \mathbf{r}\right)  $ is the electron density for the total system;
and $\delta n\left(  \mathbf{r}\right)  $ is their difference. The quantity
$G_{E}^{0}$ is the Green's function for the bare electrodes. The wavefunctions
that achieve self-consistency in the DFT framework in plane wave basis are
applied to calculate the electric current, and the electron's thermal current.

These right- and left-moving wave functions weighed with the Fermi-Dirac
distribution function according to their energies and temperatures are applied
to calculate the electric current as%
\begin{widetext}
\begin{equation}
I(\mu_{L},T_{L};\mu_{R},T_{R})=\frac{e\hbar}{mi}\int dE\int d\mathbf{r}%
_{\mathbf{\perp}}\int d\mathbf{K}_{\parallel}\left[  f_{E}(\mu_{R}%
,T_{R})I_{E{E},\mathrm{\mathbf{K}}_{||}}^{RR}(\mathbf{r})-f_{E}(\mu_{L}%
,T_{L})I_{E{E},\mathrm{\mathbf{K}}_{||}}^{LL}(\mathbf{r})\right]  , \label{I2}%
\end{equation}
\end{widetext}
where $I_{E{E}^{\prime},\mathrm{\mathbf{K}}_{||}}^{RR(LL)}(\mathbf{r})=\left[
\Psi_{E,\mathrm{\mathbf{K}}_{||}}^{R(L)}(\mathbf{r})\right]  ^{\ast}\nabla
\Psi_{{E}^{\prime},\mathrm{\mathbf{K}}_{||}}^{R(L)}(\mathbf{r})-\nabla\left[
\Psi_{E,\mathrm{\mathbf{K}}_{||}}^{R(L)}(\mathbf{r})\right]  ^{\ast}\Psi
_{{E}^{\prime},\mathrm{\mathbf{K}}_{||}}^{R(L)}(\mathbf{r})$ and
$d\mathbf{r}_{\mathbf{\perp}}$ represents an element of the electrode surface.
Here, we assume that the left and right electrodes are independent electron
reservoirs, with the population of the electron described by the Fermi-Dirac
distribution function, $f_{E}(\mu_{L(R)},T_{L(R)})=1/\{\exp[\left(
E-\mu_{L(R)}\right)  /(k_{B}T_{L(R)})]+1\}$, where $\mu_{L(R)}$ and $T_{L(R)}$
are the chemical potential and the temperature in the left (right) electrode,
respectively. More detailed descriptions of theory can be found in Refs.~[\citenum{Lang,DiVentra2,Chen}].

The above expression can be cast in a Landauer-B\"{u}ttiker formalism:%
\begin{equation}
I(\mu_{L},T_{L};\mu_{R},T_{R})=\frac{2e}{h}\int dE\tau(E)[f_{E}(\mu_{R}%
,T_{R})-f_{E}(\mu_{L},T_{L})], \label{Landau-Buttiker}%
\end{equation}
where $\tau(E)=\tau^{R}(E)=\tau^{L}(E)$ is a direct consequence of the
time-reversal symmetry, and $\tau^{R(L)}(E)$ is the transmission function of
the electrons with energy $E$ incident from the right (left) electrode,
\begin{equation}
\tau^{R(L)}(E)=\frac{\pi\hbar^{2}}{mi}\int{d\mathbf{r}}_{\perp}{\int
{d\mathrm{\mathbf{K}}_{||}}}I_{EE,\mathrm{\mathbf{K}}_{||}}^{RR(LL)}%
(\mathrm{\mathbf{r}}). \label{tau}%
\end{equation}
There is an analog between the electric current and electron's thermal
current. The flow of electrons can also transport energy. The electron's
thermal current, defined as the rate at which thermal energy flows from the
right (into the left) electrode, is
\begin{widetext}
\begin{equation}
J_{el}^{R(L)}(\mu_{L},T_{L};\mu_{R},T_{R})=\frac{2}{h}\int dE(E-\mu
_{R(L)})\tau(E)[f_{E}(\mu_{R},T_{R})-f_{E}(\mu_{L},T_{L})]. \label{JJ}%
\end{equation}
\end{widetext}
\subsection{Theory of Thermoelectricity}
Here, we present the theory to calculate the zero-bias electric
conductance, Seebeck coefficient, and electron's thermal conductance for a
thermoelectric nanojunction in terms of the effective single-particle
wave-functions obtained self-consistently within the static density-functional
theory in a truly atomic-scale junction.

We assume that the left and right electrodes serve as independent temperature
reservoirs. The population of electrons in the left (right) electrode is
described by the Fermi-Dirac distribution function, $f_{E}(\mu_{L(R)}%
,T_{L(R)})=1/\{\exp[\left(  E-\mu_{L(R)}\right)  /(k_{B}T_{L(R)})]+1\}$, where
the chemical potentials are $\mu_{R}=\mu_{L}=\mu$ (i.e., at zero external
bias). Let us now consider an extra infinitesimal current induced by an
additional infinitesimal temperature ($dT$) and voltage ($dV$) across the
junctions in a open circuit. The current induced by $dT$ and $dV$ are
$(dI)_{T}=I(\mu,T;\mu,T+dT)$ and $(dI)_{V}=I(\mu,T;\mu+edV,T)$, respectively,
where $I(\mu,T;\mu,T+dT)$ and $I(\mu,T;\mu+edV,T)$ are given by
Eq.~(\ref{Landau-Buttiker}). Suppose that the current cannot actually flow in
an open circuit, thus, $(dI)_{T}$ counterbalances $(dI)_{V}$. In other words,
the extra net current is zero,%
\begin{equation}
dI=I(\mu,T;\mu+edV,T+dT)\approx(dI)_{T}+(dI)_{V}=0.\text{ } \label{deltaI}%
\end{equation}
The Fermi-Dirac distribution function in Eq.~(\ref{deltaI}) can be expanded up
to the first order in $dT$ and $dV$, and we obtain the Seebeck coefficient
(defined by $S=dV/dT$),
\begin{equation}
S(\mu,T)=-\frac{1}{eT}\frac{K_{1}(\mu,T)}{K_{0}(\mu,T)}, \label{S}%
\end{equation}
where%
\begin{equation}
K_{n}(\mu,T)=-\int dE\tau(E)\left(  E-\mu\right)  ^{n}\frac{\partial f_{E}%
(\mu,T)}{\partial E}. \label{KKn}%
\end{equation}
The Seebeck coefficient in the low-temperature regime can be obtained by
expanding $K_{n}(\mu,T)$ to the lowest order in temperatures through the
Sommerfeld expansion, that is, $K_{0}\approx\tau(\mu)$, $K_{1}\approx
\lbrack\pi^{2}k_{B}^{2}\tau^{\prime}(\mu)/3]T^{2}$, and $K_{2}\approx
\lbrack\pi^{2}k_{B}^{2}\tau(\mu)/3]T^{2}$. The Seebeck coefficient up to the
lowest order in temperature is%
\begin{equation}
S\approx\alpha T, \label{SlowT}%
\end{equation}
where $\alpha=-\pi^{2}k_{B}^{2}\frac{\partial\tau(\mu)}{\partial E}/\left(
3e\tau(\mu)\right)  $. The Seebeck coefficient is positive (negative) when the
slope of transmission function is negative (positive), which is closely
related to the transmission function near the chemical potentials.

The electron's thermal current is the energy current carried by electrons
traveling between electrodes driven by $dT$ and $dV$. Analogous to the extra
current given by Eq.~(\ref{deltaI}), the extra electron's thermal current is
\begin{equation}
dJ_{el}=(dJ_{el})_{T}+(dJ_{el})_{V}. \label{extrJel1}%
\end{equation}
where $(dJ_{el})_{T}=J_{el}(\mu,T;\mu,T+dT)$ and $(dJ_{el})_{V}=J_{el}%
(\mu,T;\mu+edV,T)$ are the fractions of electron's thermal current driven by
$dT$ and $dV$, respectively. Note that $dV$ is generated by the Seebeck effect
according to the temperature difference $dT$. Both $J_{el}(\mu,T;\mu,T+dT)$
and $J_{el}(\mu,T;\mu+edV,T)$ can be calculated using Eq.~(\ref{JJ}).

Given that we define the electron's thermal conductance as $k_{el}=dJ_{el}%
/dT$, the electron's thermal conductance $k_{el}$ can decomposed into two
components:
\begin{equation}
\kappa_{el}(\mu,T)=\kappa_{el}^{T}(\mu,T)+\kappa_{el}^{V}(\mu,T),
\label{thercond}%
\end{equation}
where $\kappa_{el}^{T}=(dJ_{el})_{T}/dT$ and $\kappa_{el}^{V}=(dJ_{el}%
)_{V}/dT$. We note that $\kappa_{el}^{T}$ and $\kappa_{el}^{V}$ are the
portions of the electron's thermal conductance driven by $dT$ and $dV$,
respectively. Analogous to Eq.~(\ref{S}), $\kappa_{el}^{T}$ and $\kappa
_{el}^{V}$ can be can be expressed by Eq.~(\ref{KKn}):%
\begin{equation}
\kappa_{el}^{T}(\mu,T)=\frac{2}{h}\frac{K_{2}(\mu,T)}{T}, \label{kelT0}%
\end{equation}
and%
\begin{equation}
\kappa_{el}^{V}(\mu,T)=\frac{2e}{h}K_{1}(\mu,T)S(\mu,T). \label{kelV0}%
\end{equation}
One should note that $\kappa_{el}^{V}=0$ if the Seebeck coefficient of the
system is zero because $dV$ is zero.

In the low-temperature regime, $\kappa_{el}^{T}$ and $\kappa_{el}^{V}$ can be
expanded to the lowest order in temperatures using the Sommerfeld expansion:
\begin{equation}
\kappa_{el}^{V}\approx\beta_{V}T\text{ }^{3}\text{and }\kappa_{el}^{T}%
\approx\beta_{T}T, \label{kelTT}%
\end{equation}
where $\beta_{V}=-2\pi^{4}k_{B}^{4}[\tau^{\prime}(\mu)]^{2}/(9h\tau(\mu))$ and
$\beta_{T}=2\pi^{2}k_{B}^{2}\tau(\mu)/(3h)$. In the above expansions, we also
applied the following approximations: $K_{1}(\mu,T)\approx\lbrack\pi^{2}%
k_{B}^{2}\tau^{\prime}(\mu)/3]T^{2}$, $K_{2}(\mu,T)\approx\lbrack\pi^{2}%
k_{B}^{2}\tau(\mu)/3]T^{2}$, and Eq.~(\ref{SlowT}). In the low-temperature
regime, $\kappa_{el}^{T}$ dominates the electron's thermal current. Thus
$\kappa_{el}\approx\kappa_{el}^{T}$ is linear in T, that is,
\begin{equation}
\kappa_{el}(\mu,T)\approx\beta_{T}T. \label{kel2}%
\end{equation}

At zero bias, the electric conductance can be expressed as
\begin{equation}
\sigma(T)=\frac{2e^{2}}{h}\int dEf_{E}\left(  \mu,T\right)  [1-f_{E}\left(
\mu,T\right)  ]\tau(E)/(k_{B}T). \label{cond}%
\end{equation}
In the low-temperature regime, the zero-bias conductance is usually
insensitive to temperatures if tunneling is the major transport mechanism.

Thus far, the physical quantities ($S$, $\sigma$, and $\kappa_{el}$)
previously discussed are related to the propagation of electrons. The heat
current carried by phonon may occur in a real system. The phonon's thermal
current, which is driven by the temperature difference $\Delta T,$ flows from
the hot reservoir into the cold reservoir. To determine the impact of the
phonon's thermal current on refrigeration, the weak link model is chosen to
describe it. The weak link model assumes that the nanojunction is a weak
elastic link with a given stiffness $K$ that can be evaluated from total
energy calculations or from experimental measurement~\cite{Patton}. Two metal
electrodes are regarded as the macroscopic bodies under their thermodynamic
equilibrium, and are taken as ideal thermal conductors. To the leading order
in the strength of the weak link, the phonon's thermal current ($J_{ph}$) via
elastic phonon scattering is~\cite{Patton},
\begin{equation}
J_{ph}=\frac{2\pi K^{2}}{\hbar}\int_{0}^{\infty}dEEN_{L}(E)N_{R}%
(E)[n_{L}(E)-n_{R}(E)], \label{JQ}%
\end{equation}
where $K$ is the stiffness of the weak elastic link; $N_{L(R)}(E)$ is the
spectral density of phonon states at the left (right) electrode surface which
is measurable by experiments; and $n_{L(R)}\equiv1/(e^{E/K_{B}T_{L(R)}}-1)$ is
the Bose-Einstein distribution function. In the long wave length limit, the
spectral density of surface phonon is given by $N_{L(R)}(E)\approx CE$. The
phonon's thermal conductance defined as $\kappa_{ph}=\Delta J_{ph}/\Delta T$
as $\Delta T\rightarrow0$, is
\begin{equation}
\kappa_{ph}=[J_{ph}(T_{L}+\frac{\Delta T}{2},T_{R}-\frac{\Delta T}{2}%
)-J_{ph}(T_{L},T_{R})]/\Delta T. \label{kqph}%
\end{equation}
Expanding the Bose-Einstein distribution function in $J_{ph}(T_{L}%
+\frac{\Delta T}{2},T_{R}-\frac{\Delta T}{2})$ to the first order of $\Delta
T$, we obtain the phonon's thermal conductance:
\begin{equation}
\kappa_{ph}=\frac{\pi K^{2}C^{2}}{\hbar}\int_{0}^{\infty}dEE^{3}\sum
_{i=L,R}\frac{\partial n_{i}(E)}{\partial T_{i}}. \label{kph}%
\end{equation}
When $T_{R}\approx T_{L}=T$, then Eq.~(\ref{kph}) is reduced to,
\begin{equation}
\kappa_{ph}=\frac{2\pi K^{2}C^{2}}{\hbar}\int_{0}^{\infty}dEE^{3}%
\frac{\partial n(E)}{\partial T}. \label{kph1}%
\end{equation}
where $n=1/[e^{E/(k_{B}T)}-1]$.

The efficiency of thermoelectric nano-devices is conventionally described by
the thermoelectric figure of merit $ZT$, which depends on the following
physical factors: Seebeck coefficient $(S)$, electric conductance $(\sigma)$,
electron's thermal conductance $(\kappa_{el})$, and phonon's thermal
conductance $(\kappa_{ph})$. $ZT$ is defined as
\begin{equation}
ZT=\frac{S^{2}\sigma}{\kappa_{el}+\kappa_{ph}}T, \label{zt}%
\end{equation}
where $S$, $\sigma$, $\kappa_{el}$, and $\kappa_{ph}$ can be numerically
calculated using Eqs.~(\ref{S}),~(\ref{cond}),~(\ref{thercond}),
and~(\ref{kph1}), respectively. When $ZT$ tends to infinity, the
thermoelectric efficiency of nanojunctions will reach Carnot efficiency.
\subsection{Theory of Thermoelectric Refrigerator}
\label{II.C}
Now, we present the theory of thermoelectric cooling device
at atomic scale including effect of the phonon's thermal current and
local heating. We assume that the left (right) electrode serves as the hot
(cold) temperature reservoir with temperature $T_{R}=T_{C}$ ($T_{L}%
=T_{H}=T_{C}+\Delta T$) and the phonon's population is described by
Bose-Einstein distribution function. We consider the nanojunction connecting
to an external battery with bias $V_{B}=(\mu_{R}-\mu_{L})/e$, which drives the
electrons flowing from the right- to left-electrodes. The thermal current
carried by electrons traveling between two electrodes is given by Eq.~(\ref{JJ}).
It should be noted that
\begin{widetext}
\begin{equation}
J_{el}^{R}(\mu_{L},T_{L};\mu_{R},T_{R})+I(\mu_{L},T_{L};\mu_{R},T_{R}%
)V_{B}=J_{el}^{L}(\mu_{L},T_{L};\mu_{R},T_{R}), \label{energyconversation}%
\end{equation}
\end{widetext}
which $I(\mu_{L},T_{L};\mu_{R},T_{R})$ is the current given by
Eq.~(\ref{Landau-Buttiker}). Equation~(\ref{energyconversation}) states that
the energy is conserved: it consumes electric energy to take heat from the
right (cold) into left (hot) reservoir. Thus, the thermoelectric junction can
be regarded as an electronic cooling device when $J_{el}^{R}>0$, which states
that the thermoelectric junction is capable of removing heat from the cold reservoir.

A measure of a refrigerator's performance is the ratio of the rate of heat
removed from the cold reservoir to the electric power done on the system. The
ratio is called the coefficient of performance (COP):
\begin{equation}
\eta_{el}=\frac{J_{el}^{R}}{IV_{B}}=\frac{J_{el}^{R}}{\left\vert J_{el}%
^{R}-J_{el}^{L}\right\vert }, \label{eta1}%
\end{equation}
where $V_{B}$ is the bias applied across the nanojunction driving the
thermoelectric refrigerator, $I$ is the electric current, and $J_{el}^{R}$ is
the rate of thermal energy removed from the cold reservoir. The thermoelectric
junction as a nano-refrigerator is working when $J_{el}^{R}>0$ (and thus
$\eta_{el}>0$).
\subsubsection{Properties of Thermoelectric Refrigerator in the
Absence of the Phonon's Thermal Current}
In the following, we develop an analytical theory to gain insight into the
fundamentals of the cooling effect in the thermoelectric nanojunction. We
apply the Sommerfeld expansion to $J_{el}^{R}(\mu_{L},T_{L};\mu_{R},T_{R})$
and obtain,%
\begin{align}
J_{el}^{R}  &  \approx\frac{\pi^{2}k_{B}^{2}}{3h}\left[  \tau(\mu_{L}%
)(T_{R}^{2}-T_{L}^{2})+\tau^{\prime}(\mu_{L})(\mu_{R}-\mu_{L})(T_{R}^{2}%
+T_{L}^{2})\right] \nonumber\\
&  -\frac{1}{h}\left[  \tau(\mu_{L})(\mu_{R}-\mu_{L})^{2}\right]  -\frac
{1}{3h}\left[  \tau^{\prime}(\mu_{L})(\mu_{R}-\mu_{L})^{3}\right]  ,
\label{JRe1}%
\end{align}
where we use the following relations: $\tau(E)\approx$ $\tau(\mu_{L}%
)+\tau^{\prime}(\mu_{L})(E-\mu_{L})$, $\tau(\mu_{R})\approx$ $\tau(\mu
_{L})+\tau^{\prime}(\mu_{L})(\mu_{R}-\mu_{L})$, $T_{L}=T_{R}+\Delta T$,
$\int_{\mu_{L}}^{\mu_{R}}(E-\mu_{R})dE=-\frac{1}{2}(\mu_{R}-\mu_{L})^{2}$,
$\int_{\mu_{L}}^{\mu_{R}}(E-\mu_{L})(E-\mu_{R})dE=-\frac{1}{6}(\mu_{R}-\mu
_{L})^{3}$, and $\mu_{R}-\mu_{L}=eV_{B}$. When $\Delta T\ll T_{R}$,
Eq.~(\ref{JRe1}) can be expressed as a polynomial of $V_{B}$:%
\begin{equation}
J_{el}^{R}=-a+bV_{B}-cV_{B}^{2}-dV_{B}^{3}, \label{JRpoly}%
\end{equation}
where $a=2\pi^{2}k_{B}^{2}\tau(\mu)T_{R}\Delta T/(3h)$, $b=2e\pi^{2}k_{B}%
^{2}\tau^{\prime}(\mu)T_{R}^{2}/(3h)$, $c=e^{2}\tau(\mu)/h$, and $d=e^{3}%
\tau^{\prime}(\mu)/(3h)$. The above equation is convenient for analytical
exploration of the properties of thermoelectric nano-refrigerators.

A thermoelectric nano-refrigerator functions only when the maximum value of
the electron's thermal current is positive, that is, $\left(  J_{el}%
^{R}\right)  _{\max}>0$. In a small bias regime, the term $V_{B}^{3}$ can be
neglected in Eq.~(\ref{JRpoly}), i.e., $J_{el}^{R}\approx-a+bV_{B}-cV_{B}^{2}%
$. In this case, the working condition of nano-refrigerator is given by
$\left(  J_{el}^{R}\right)  _{\max}=(-4ac+b^{2})/(4c)>0$, which yields the
criterion for the existence of electronic cooling
\begin{equation}
-S>\sqrt{\frac{2\pi^{2}k_{B}^{2}}{3e^{2}}\left(  \frac{\Delta T}{T_{R}%
}\right)  }, \label{criterion}%
\end{equation}
where $S=-\frac{\pi^{2}k_{B}^{2}}{3e}\frac{\tau^{\prime}(\mu)}{\tau(\mu)}%
T_{R}$ is the Seebeck coefficient of the nanoscale junction~\cite{Liu}.

For a given $T_{R}$ and a given temperature difference $\Delta T$, $J_{el}%
^{R}(V_{B})$ is a function of bias $V_{B}$. We observe that $J_{el}^{R}$ is
negative at $V_{B}=0$, and thus a lower limit of bias (denoted as
$V_{el}^{th,lower}$) is needed. The lower threshold bias is defined as the
smallest positive solution of $J_{el}^{R}(V_{B})=0$. For $V_{B}<V_{el}%
^{th,lower}$, $J_{el}^{R}<0$ and thermoelectric cooling effect does not exist.
Moreover, it is observed that there is an upper bound of bias for the
operation of thermoelectric nano-refrigerator. To show this, we keep the terms
up to $O(V_{B}^{2})$ in Eq.~(\ref{JRpoly}), i.e., $J_{el}^{R}(V_{B}%
)\approx-a+bV_{B}-cV_{B}^{2}$. In this simple case, the lower threshold bias
$V_{el}^{th,lower}$ and the upper threshold bias $V_{el}^{th,upper}$ is
derived from $J_{el}^{R}(V_{B})=0$, from which we obtain the lower and upper
bounds of the working biases,%
\begin{equation}
V_{el}^{th,lower}\approx-\frac{\pi^{2}k_{B}^{2}\Delta T}{3e^{2}ST_{R}},
\label{Vth_el2}%
\end{equation}
and%
\begin{equation}
V_{el}^{th,upper}\approx2ST_{R}^{2}-V_{th}^{el}\approx2ST_{R}^{2},
\label{Vupper_el}%
\end{equation}
where we have assumed $\Delta T\ll T_{R}$. Eqs.~(\ref{Vth_el2}) and
(\ref{Vupper_el}) impose a constraint for applied biases which allow the
thermoelectric refrigeration. The nano-refrigerator functions only when
$V_{el}^{th,lower}<V_{B}<V_{el}^{th,upper}$. Equation~(\ref{Vth_el2}) shows
that the lower threshold bias $V_{el}^{th,lower}$ slightly decreases as
$T_{R}$ and $\Delta T$ increases. Eq.~(\ref{Vupper_el}) predicts that the
upper threshold bias $V_{el}^{th,upper}$ increases as $T_{R}$ increases.

We note that $J_{el}^{R}$ is a function of bias $V_{B}$, $T_{R}$, and $\Delta
T$. For a given bias $V_{B}$ and a given temperature difference $\Delta T$,
$J_{el}^{R}$ is a function of $T_{R}$. We observe that there is a lower limit
of temperatures when the thermoelectric refrigerator is working. The onset
temperature for refrigeration effect, denoted as $T_{el}^{OP}$, is defined by
$J_{el}^{R}(T_{el}^{OP})=0$. The nano-refrigerator is not functioning
($J_{el}^{R}(T_{R})<0$) when $T_{R}<T_{el}^{OP}$. Especially, $J_{el}%
^{R}(T_{R})$ can be expressed as a polynomial of $T_{R}$, derived from
Eq.~(\ref{JRe1}). If the Seebeck coefficient is sufficiently large and we
neglect the terms higher than $O(T_{R}^{2})$, the threshold operation
temperature $T_{el}^{OP}$ can be calculated analytically by solving the
polynomial $J_{el}^{R}(T_{R})=0$, which gives:
\begin{equation}
T_{el}^{OP}\approx\frac{1}{2}[\alpha/V_{B}+\sqrt{(\alpha/V_{B})^{2}+\beta
V_{B}}], \label{Top1}%
\end{equation}
where $\alpha=\frac{\tau(\mu)\Delta T}{\tau^{\prime}(\mu)e}$ and $\beta
=\frac{6e\tau(\mu)}{\pi^{2}k_{B}^{2}\tau^{\prime}(\mu)}$. Equation~(\ref{Top1}%
) shows that $T_{el}^{OP}$ increases as $\Delta T$ and $V_{B}$ increase,
respectively. When $\Delta T=0$, $\alpha=0$ and Eq.~(\ref{Top1}) approaches to%
\begin{equation}
\left(  T_{el}^{OP}\right)  _{\min}\approx\sqrt{\frac{V_{B}T_{R}}{-2S}}.
\label{Top_min}%
\end{equation}
Equation~(\ref{Top_min}) shows that $T_{el}^{OP}$ approaches to the lower
limit $\left(  T_{el}^{OP}\right)  _{\min}$ as $\Delta T\rightarrow0$, where
$\left(  T_{el}^{OP}\right)  _{\min}$, increases as $V_{B}\times T_{R}$ increases.

We now turn to investigating the COP of the thermoelectric cooling device. When
$V_{B}$ is small, $J_{el}^{L}-J_{el}^{R}=IV_{B}\approx\sigma V_{B}^{2}$ . In
this case, the COP is in the following form,
\begin{equation}
\eta_{el}\approx\frac{J_{el}^{R}}{\sigma V_{B}^{2}}, \label{COP2}%
\end{equation}
where $J_{el}^{R}(V_{B})$ is a polynomial of $V_{B}$ given by Eq.~(\ref{eta1}).
The applied biases considered in this study are small, hence we
can consider $J_{el}^{R}(V_{B})$ [given by Eq.~(\ref{JRpoly})] up to
$O(V_{B})$. Thus, Eq.~(\ref{COP2}) takes the form,%
\begin{equation}
\eta_{el}\approx\frac{-a+bV_{B}}{\sigma V_{B}^{2}}. \label{COP4}%
\end{equation}
The maximum value of COP (denoted as $\eta_{\max}^{el}$) occurs at
$\eta^{\prime}(V_{\max}^{\eta})=0$, which gives%
\begin{equation}
\eta_{\max}^{el}\approx\frac{3e^{2}S^{2}T_{R}}{4\pi^{2}k_{B}^{2}\Delta
T}-\frac{1}{2}, \label{eta_el_max2}%
\end{equation}
where the maximum value of $\eta_{el}$ occurs at bias $V_{B}=V_{\max}%
^{\eta_{el}}$, where%
\begin{equation}
V_{\max}^{\eta_{el}}\approx2V_{el}^{th,lower}, \label{V_eta_max}%
\end{equation}
where $V_{el}^{th,lower}$ is given by Eq.~(\ref{Vth_el2}).
\subsubsection{Effect of the Phonon's Thermal Current}
Phonon's thermal current flows from the hot to cold reservoir. It is an
adverse effect to thermoelectric refrigeration because it heats up cold
electrodes. To realize the impact of this adverse effect to refrigeration, we
consider the weak-link model suitable for describing the heat transport for
two thermal reservoirs connected by a weak elastic link~\cite{Patton}. In the
low-temperature regime ($T\ll T_{D}$, where $T_{D}=394$~K is the Debye
temperature for Al), Eq.~(\ref{kph}) can be expanded up to the lowest order in
temperatures,%
\begin{equation}
J_{ph}^{R}=\lambda(T_{R}^{4}-T_{L}^{4}), \label{simplifiedJph}%
\end{equation}
where $\lambda\approx2\pi^{5}K^{2}C^{2}k_{B}^{4}/(15\hbar)$, where $K$ is the
stiffness of the nano-structured object bridging the metal electrodes, and $C$
is the slope of the surface phonon's dispersion function in long wavelength
limit. The effect of the phonon's thermal current on refrigeration is
described by a single parameter $\lambda$, which is determined by $K$ and $C$.
The simplified weak link model allows us to develop an analytical
theory to investigate the effect of the phonon's thermal current on
refrigeration using a single parameter $\lambda$. The influence of the
strength of the phonon's thermal current on thermoelectric refrigeration
becomes transparent.

To determine the impact of the phonon's thermal current ($J_{ph}^{R}$), we
repeat similar discussions in the previous subsection for the thermoelectric
nano-refrigerators. Correspondingly, the COP of the thermoelectric
nano-refrigerator becomes
\begin{equation}
\eta^{el+ph}=\frac{J_{el+ph}^{R}}{IV_{B}}, \label{etaelph}%
\end{equation}
where $J_{el+ph}^{R}=J_{el}^{R}+J_{ph}^{R}$ is the combined thermal current
including the phonon's thermal current.

We assume that $\Delta T\ll T_{R}$ and Eq.~(\ref{simplifiedJph}) is
approximated as,%
\begin{equation}
J_{ph}^{R}\approx-4\lambda T_{R}^{3}\Delta T. \label{JRphdt}%
\end{equation}
Using Eqs.~(\ref{JRpoly}) and (\ref{JRphdt}), the combined thermal current can
be expressed in terms of a polynomial of $V_{B}$ similar to Eq.~(\ref{JRpoly}%
),%
\begin{equation}
J_{el+ph}^{R}=-a_{el+ph}+bV_{B}-cV_{B}^{2}-dV_{B}^{3}, \label{JRpolyph}%
\end{equation}
where $a_{el+ph}=2\pi^{2}k_{B}^{2}\tau(\mu)T_{R}\Delta T/(3h)+4\lambda
T_{R}^{3}\Delta T$; the coefficients $b$, $c$, and $d$ remain the same as
those in Eq.~(\ref{JRpoly}).

An analogy can be drawn here between Eqs.~(\ref{JRpolyph}) and (\ref{JRpoly}),
and likewise between Eqs.~(\ref{etaelph}) and (\ref{eta1}). Consequently, the
equations in the previous section can be easily replicated here. In the
presence of the phonon's thermal current, the working condition of the
nano-refrigerator becomes,%
\begin{equation}
-S>\sqrt{\left(  \frac{2\pi^{2}k_{B}^{2}}{3e^{2}}+\lambda\frac{8 T_{R}^{2}}{\sigma
}\right)  \left(  \frac{\Delta T}{T_{R}}\right)  }, \label{criterion_ph}%
\end{equation}
where $\lambda$ represents the strength of the phonon's thermal current, and
$\sigma$ is the electric conductance of the nanojunction. As the phonon's
thermal current vanishes ($\lambda=0$), Eq.~(\ref{criterion_ph}) restores
Eq.~(\ref{criterion}). A sufficiently large phonon's thermal current has a large
value of $\lambda$, which is likely to break down the inequality described in
Eq.~(\ref{criterion_ph}) and ruin refrigeration capability. Equation
(\ref{criterion_ph}) shows that the phonon's thermal current is an adverse
effect to thermoelectric refrigeration.

Similar to Sec.~II.~C, the thermoelectric nano-refrigeration
works in a small range of biases. For a given $T_{R}$ and a given temperature
difference $\Delta T$, $J_{el+ph}^{R}(V_{B})$ is a function of bias $V_{B}$.
We note that $J_{el+ph}^{R}$ is negative at $V_{B}=0$; therefore, a minimum
bias (denoted as $V_{el+ph}^{th,lower}$) is needed to trigger possible
thermoelectric cooling effect. Similarly, the upper threshold bias
$V_{el+ph}^{th,upper}$ is defined as the second zero of $J_{el}^{R}(V_{B}%
)=0$. Owing to $\Delta T\ll T_{R}$ and the small values of $V_{el+ph}%
^{th,lower}$ and $V_{el+ph}^{th,upper}$, the lower threshold bias becomes%
\begin{equation}
V_{el+ph}^{th,lower}\approx V_{el}^{th,lower}+\lambda\left[  4T_{R}^{2}\Delta
T/\left(  -S\right)  \sigma\right]  , \label{Vth_elph2}%
\end{equation}
where $V_{el}^{th,lower}$ is given by Eq.~(\ref{Vth_el2}) which is analogous
to Eq.~(\ref{Vth_elph2}). The lower threshold bias $V_{el+ph}^{th,lower}$
increases as the intensity of the phonon's thermal current ($\lambda$)
increases. Similarly, the upper threshold bias $V_{el+ph}^{th,upper}$ is given by,%
\begin{equation}
V_{el+ph}^{th,upper}\approx V_{el}^{th,upper}-\lambda\left[  4T_{R}^{2}\Delta
T/\left(  -S\right)  \sigma\right]  , \label{Vupper_elph2}%
\end{equation}
where $V_{el}^{th,upper}$ is given by Eq.~(\ref{Vupper_el}). The upper
threshold bias $V_{el+ph}^{th,upper}$ decreases as the intensity of the
phonon's thermal current ($\lambda$) increases. The range of working biases
shrinks by the phonon's thermal current, as given by Eqs.~(\ref{Vupper_elph2})
and (\ref{Vupper_elph2}).

For a fixed $\Delta T$, the the functioning of the thermoelectric
refrigerator is restricted to a range of temperatures between $T_{low}^{OP}$
and $T_{high}^{OP}$ obtained from Eq.~(\ref{criterion_ph}). The lower and upper
bounds of the operation temperatures are given by,%
\begin{equation}
T_{low,el+ph}^{OP}=\frac{S^{2}+\sqrt{S^{2}-\lambda(\frac{64\pi^{2}k_{B}%
^{2}\Delta T}{3e\sigma})}}{\lambda(16\Delta T/\sigma)},\label{Topphlow}%
\end{equation}
and%
\begin{equation}
T_{high,el+ph}^{OP}=\frac{S^{2}-\sqrt{S^{2}-\lambda(\frac{64\pi^{2}k_{B}%
^{2}\Delta T}{3e\sigma})}}{\lambda(16\Delta T/\sigma)}.\label{Topphhigh}%
\end{equation}

Finally, we investigate the maximum value of the COP for a fixed $T_{R}$ and a
fixed temperature difference $\Delta T$. The optimized COP considering the
phonon's thermal current is given by%
\begin{equation}
\eta_{el+ph}^{\max}\approx\left[  \left(  \eta_{el}^{\max}\right)
^{-1}+\lambda\left(  \frac{16T_{R}\Delta T}{S^{2}\sigma}\right)  \right]
^{-1}, \label{eta_elph_max2}%
\end{equation}
where $\eta_{el}^{\max}$ is the upper limit of COP given by
Eq.~(\ref{eta_el_max2}) which is analogous to Eq.~(\ref{eta_elph_max2}).
\subsubsection{Effect of Local Heating}
\begin{figure}[ptb]
\includegraphics[width=7.5cm]{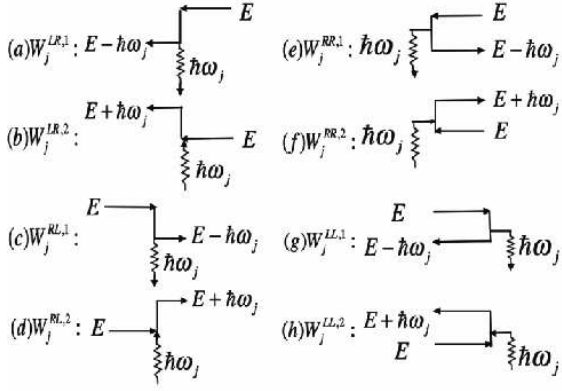}
\caption{
Feymann diagrams of the first-order electron-vibration scattering processes
considered in this study.
}%
\label{Fig2}%
\end{figure}
Following the work of Chen, Zwolak, and Di Ventra~\cite{chenheating}, the
many-body Hamiltonian of the system, which considers the vibration of the
atom/molecule bridging the electrodes, is
\begin{equation}
H=H_{el}+H_{vib}+H_{el-vib}, \label{hamil}%
\end{equation}
where $H_{el}$ is the electronic part of the Hamiltonian under adiabatic
approximations and $H_{vib}$ is the ionic part of the Hamiltonian considered
in normal coordinates,
\begin{equation}
H_{vib}=\frac{1}{2}\sum_{j\in vib}\dot{q}_{j}^{2}+\frac{1}{2}\sum_{j\in
vib}\omega_{j}^{2}q_{j}^{2}, \label{Hvib}%
\end{equation}
where $\left\{  \omega_{j}\right\}  _{j=1,3N}$ are the normal mode
frequencies; $\left\{  q_{j}\right\}  _{j=1,3N}$ are the normal coordinates
which are related to the Cartesian coordinates by,%
\begin{equation}
(\mathbf{Q}_{i})_{\mu}=\sum_{j\in vib}A_{i\mu,j}q_{j} \label{normalcoor}%
\end{equation}
where $\mathbf{Q}_{i}=\mathbf{R}_{i}-\mathbf{R}_{i}^{0}$ is a small deviation
of the $i$-th ion from its equilibrium position $\mathbf{R}_{i}^{0}$ and $\mu$
\{$=x$, $y$, $z$\} denotes the \{$x$, $y$, $z$\}-component; $A_{i\mu,j}$ is a
transformation between normal and Cartesian coordinates satisfying the
canonical transformation: $%
{\displaystyle\sum\nolimits_{i,\mu}}
A_{i\mu,j}A_{i\mu,j^{\prime}}=\delta_{j,j^{\prime}}$. $H_{el-vib}$ is a part
of the Hamiltonian for electron-vibration interactions which has the form of,%
\begin{align}
H_{el-vib}  &  =\sum_{\alpha,\beta,E_{1},E_{2},j}\left(  \sum_{i,\mu}%
\sqrt{\frac{\hbar}{2M_{i}\omega_{j}}}A_{i\mu,j}J_{E_{1},E_{2}}^{i\mu
,\alpha\beta}\right) \nonumber\\
&  \cdot a_{E_{1}}^{\alpha\dag}a_{E_{2}}^{\beta}(b_{j}+b_{j}^{\dag}),
\label{elph}%
\end{align}
where $\alpha,\beta=\{L,R\}$; $M_{i}$ is the mass of the $i$-th atom; ;
$b_{j}(b_{j}^{\dagger})$ are the phonon annihilation (creation) operators for
the $j$-th vibrational mode of nanoscale junctions and they satisfy the
commutation relation $[b_{j},b_{j^{\prime}}^{\dagger}]=\delta_{j,j^{\prime}}$,
and $\{a_{E}^{L(R)}\dagger,a_{E}^{L(R)}\}$ are the creation and annihilation
operators respectively for incident electrons with energy $E$ from the left
(right) electrode. They satisfy the usual anticommutation relation,
$\{a_{E_{1}}^{\alpha},a_{E_{2}}^{\beta\dagger}\}=\delta_{\alpha\beta}%
\delta(E_{1}-E_{2})$; the coupling constant $J_{E_{1},E_{2}}^{i\mu,\alpha
\beta}$ between electrons and the vibration of the $i$-th atom in $\mu$ ($=x$,
$y$, $z$) component can be calculated as,%
\begin{equation}
J_{E_{1},E_{2}}^{i\mu,\alpha\beta}=\int d\mathbf{r}\int d\mathbf{K}%
_{\parallel}[\Psi_{E_{1}\mathbf{K}_{\parallel}}^{\alpha}(\mathbf{r})]^{\ast
}[\partial_{\mu}V^{ps}(\mathbf{r},\mathbf{R}_{i})\Psi_{E_{2}\mathbf{K}%
_{\parallel}}^{\beta}(\mathbf{r})], \label{couplingJ}%
\end{equation}
where $V^{ps}(\mathbf{r},\mathbf{R}_{i})$ is the pseudopotential representing
the interaction between electrons and the $i$-th ion; $\Psi_{E\mathbf{K}%
_{\parallel}}^{\alpha(=L,R)}(\mathbf{r})$ stands for the effective
single-particle wave function of the entire system corresponding to incident
electrons propagated from the left (right) electrode. These wave functions are
calculated iteratively until convergence and self-consistency are achieved in
the framework of DFT combined with the Lippmann-Schwinger equation~\cite{Lang}.

We now use the first-order time-independent perturbation theory to approximate
the wave function. The unperturbed system where electron-phonon scattering is
absent can be described by $|\Psi_{E}^{L(R)};n_{j}\rangle=|\Psi_{E}%
^{L(R)}\rangle\otimes|n_{j}\rangle$ where $|n_{j}\rangle$ is the phonon state
of the $j$-th normal mode~\cite{chenheating}. In Fig.~\ref{Fig2}, we display
eight different electron-phonon scattering processes when electrons tunnel
through nanoscale junctions.

Since electron-vibration interaction is directly related to junction heating,
we also include local heating in our Seebeck coefficient calculations. Details
of the theory of local heating in nanoscale structures can be found in
Ref.\citenum{chenheating}. The power absorbed and emitted by electrons incident
from the $\beta=\{L,R\}$ electrode to the $\alpha=\{L,R\}$ electrode via a
vibrational mode $j$ is denoted by $W_{j}^{\alpha\beta,k}$. The total thermal
power $P$ generated in the junction can be written as the sum over all
vibrational modes of eight scattering processes shown in Fig.~\ref{Fig2} ,
\begin{align}
P  &  =\sum_{j\in vib}(W_{j}^{RR,2}+W_{j}^{RL,2}+W_{j}^{LR,2}+W_{j}%
^{LL,2}\nonumber\\
&  -W_{j}^{RR,1}-W_{j}^{RL,1}-W_{j}^{LR,1}-W_{j}^{LL,1}), \label{power}%
\end{align}
where $W_{j}^{\alpha\beta,k}$ are calculated from the Fermi golden rule,%
\begin{align}
W_{j}^{\alpha\beta,k}  &  =2\pi\hbar(\delta_{k,2}+\left\langle n_{j}%
\right\rangle )\int dE\left\vert \sum_{i,\mu}A_{i\mu,j}J_{E\pm\hbar\omega
_{j},E}^{i\mu,\alpha\beta}\right\vert ^{2}\nonumber\\
&  \cdot f_{E}\left(  \mu_{\alpha},T_{\alpha}\right)  [1-f_{E\pm\hbar
\omega_{j}}\left(  \mu_{\beta},T_{\beta}\right)  ]D_{E\pm\hbar\omega_{j}%
}^{\alpha}D_{E}^{\beta}, \label{Golden}%
\end{align}
where $\alpha,\beta=\{L,R\}$ and $\delta_{k,2}$ is the Kronecker delta and
$k=1$ $(2)$ corresponding to relaxation (excitation) of the vibrational modes;
$D^{\alpha}$\ is partial density of states corresponding to $\Psi_{E}^{\alpha
}$; the ensemble average of occupation number of the $j$-th vibrational mode
is $\left\langle n_{j}\right\rangle =1/\{\exp[\hbar\omega_{j}/(k_{B}%
T_{w})]-1\}$ , where $T_{w}$ is the effective wire temperature. The majority
of the heat generation in the central region of the atomic junction is
transferred to electrodes. We estimate the rate of heat dissipation using the
simplified weak link model. We assume that the rate of thermal current from the
junction with temperature $T_{w}$ dissipated to the left electrode with
temperature $T_{L}$ is equivalent to the thermal current of a weak thermal
link between reservoirs with temperature $T_{L}$ and $T_{R}=2T_{w}-T_{L}$. The
rate of heat generation transferred to the left electrode is, therefore,
approximated to,%
\begin{equation}
J_{local~heating}^{L}=\lambda\lbrack(2T_{w}-T_{L})^{4}-T_{L}%
^{4}].\label{JLheating}%
\end{equation}
Similarly, the rate of heat generation transferred to the right electrode
approximately is,%
\begin{equation}
J_{local~heating}^{R}=\lambda\lbrack(2T_{w}-T_{R})^{4}-T_{R}%
^{4}].\label{JRheating}%
\end{equation}
The effective local temperature $T_{w}$ is obtained when heat generation in
the nano-structure [Eq.~(\ref{power})] and heat dissipation into the bulk
electrodes [Eqs.~(\ref{JLheating}) and (\ref{JRheating})] reach balance,%
\begin{equation}
P_{local~heating}=J_{local~heating}^{L}+J_{local\text{ }%
heating}^{R}.\label{balance}%
\end{equation}

We calculate the effective local temperature $T_{w}$ by
solving Eq.~(\ref{balance}). When considering the effect of local heating and
phonon's thermal current, the rate of the heat energy extracted from the cold
(right) reservoir is,%
\begin{equation}
J_{el+ph+heating}^{R}=J_{el+ph}^{R}-J_{local\text{ }heating}^{R}%
,\label{JR_elphheating}%
\end{equation}
and the corresponding COP of the thermoelectric nano-refrigerator becomes,
\begin{equation}
\eta_{el+ph+heating}=\frac{J_{el+ph+heating}^{R}}{IV_{B}}%
.\label{eta_localheating}%
\end{equation}
\section{Results and Discussion}
\begin{figure}[ptb]
\includegraphics[width=7.5cm]{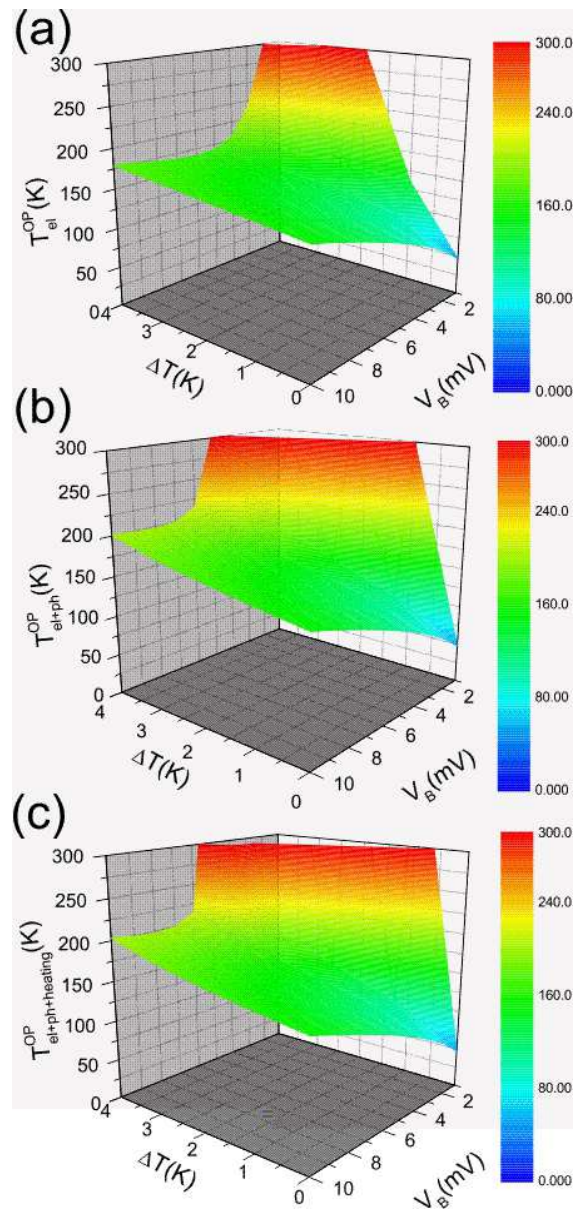}
\caption{(color online)
(a) Critical operation temperature $T_{el}^{OP}$ vs. $\Delta T$
and $V_{B}$ in the absence of the phonon's thermal current;
(b) $T_{el+ph}^{OP}$ vs. $\Delta T$ and $V_{B}$ in the presence of the
phonon's thermal current with $\lambda=10^{-17}$~Watt/K$^{4}$;
and
(c) $T_{el+ph+heating}^{OP}$ vs. $\Delta T$ and $V_{B}$ in the presence of
local heating and the phonon's thermal current
with $\lambda=10^{-17}$~Watt/K$^{4}$.}%
\label{Fig3}%
\end{figure}
\begin{figure}[ptb]
\includegraphics[width=7.5cm]{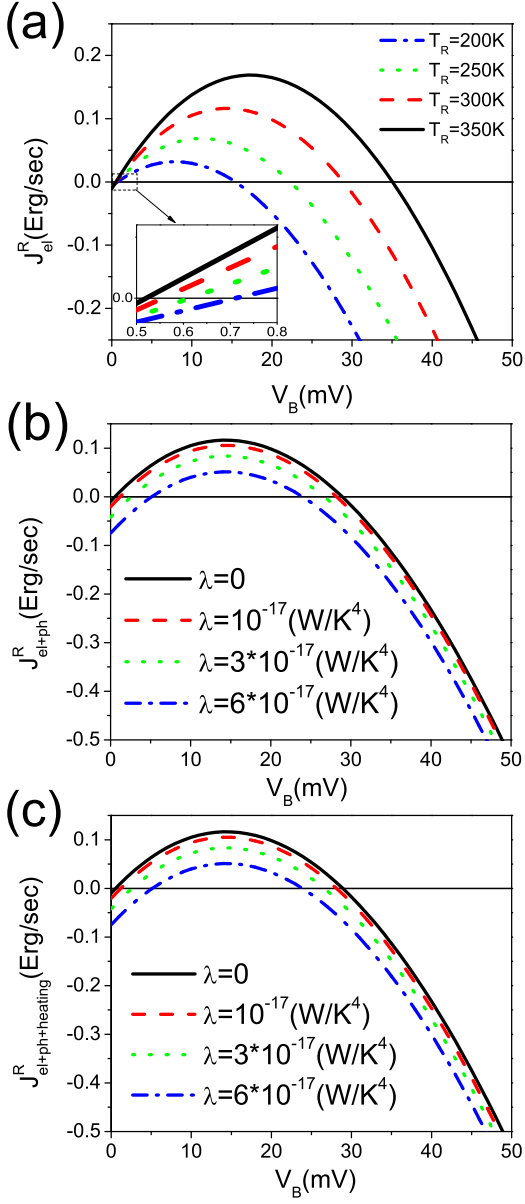}
\caption{
(color online)
(a) $J_{el}^{R}$ vs. $V_{B}$ for various values of $T_{R}$ in the absence of the
phonon's thermal current, where $\Delta T=1~$K;
(b) $J_{el+ph}^{R}$ vs. $V_{B}$ for $\lambda=0$, $1\times10^{-17}$,
$3\times10^{-17}$, and $6\times10^{-17}$~W/K$^{4}$ in the presence of
the phonon's thermal current, where $\Delta T=1$ K and $T_{R}=300$ K;
(c) $J_{el+ph+heating}^{R}$ vs. $V_{B}$ for $\lambda=0$, $1\times10^{-17}$,
$3\times10^{-17}$, and $6\times10^{-17}$~W/K$^{4}$ in the presence of
local heating and the phonon's thermal current, where $\Delta T=1$~K and $T_{R}=300$~K.
}%
\label{Fig4}%
\end{figure}
\begin{figure}[ptb]
\includegraphics[width=7.5cm]{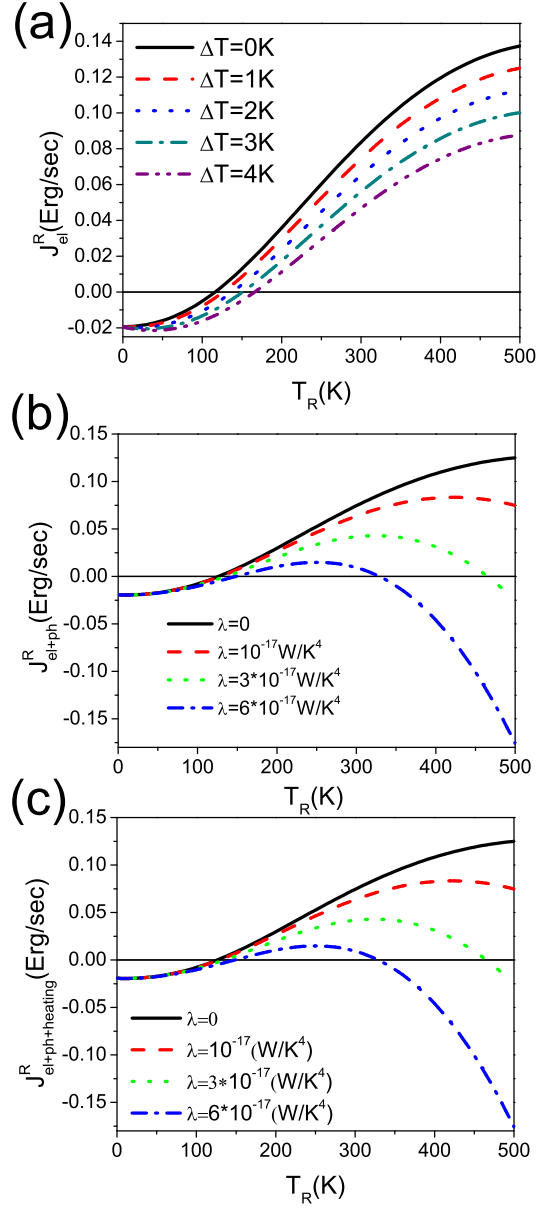}
\caption{
(color online)
(a) $J_{el}^{R}$ vs. $T_{R}$ for various values of
$\Delta T=0$, $1$, $2$, $3$, and $4$~K in the absence of the
phonon's thermal current, where $V_{B}=6$~mV;
(b) $J_{el+ph}^{R}$ vs. $T_{R}$ for $\lambda=0$, $1\times10^{-17}$,
$3\times10^{-17}$, and $6\times10^{-17}$~W/K$^{4}$ in the presence of
the phonon's thermal current, where $V_{B}=6$~mV;
(c) $J_{el+ph+heating}^{R}$ vs. $T_{R}$ for $\lambda=0$, $1\times10^{-17}$,
$3\times10^{-17}$, and $6\times10^{-17}$~W/K$^{4}$ in the presence of
local heating and the phonon's thermal current, where $V_{B}=6$~mV.
}%
\label{Fig5}%
\end{figure}
\begin{figure}[ptb]
\includegraphics[width=7.5cm]{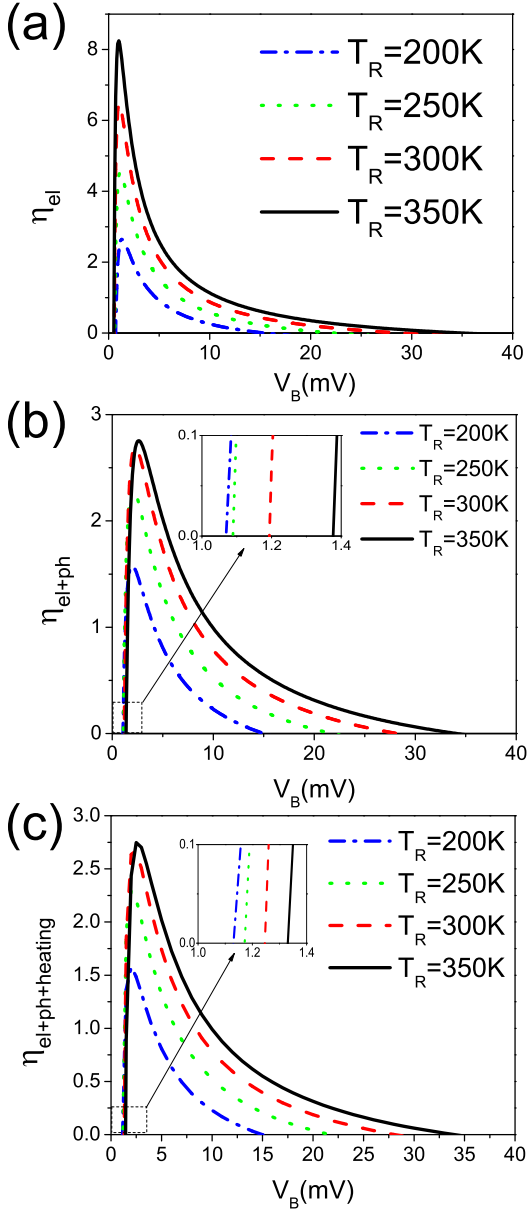}
\caption{
(color online)
(a) COP ($\eta_{el}$) vs. $V_{B}$ in the absence of the phonon's thermal
current for $T_{R}=200$, $250$, $300$, and $350$~K, where $\Delta T=1$~K;
(b) $\eta_{el+ph}$ vs. $V_{B}$ in the presence of the phonon's thermal current
for $T_{R}=200$, $250$, $300$, and $350$~K, where $\Delta T=1$~K;
(c) $\eta_{el+ph+heating}$ vs. $V_{B}$ in the presence of
local heating and the phonon's thermal current
for $T_{R}=200$, $250$, $300$, and $350$~K, where $\Delta T=1$~K.
}%
\label{Fig6}%
\end{figure}
\begin{figure}[ptb]
\includegraphics[width=7.5cm]{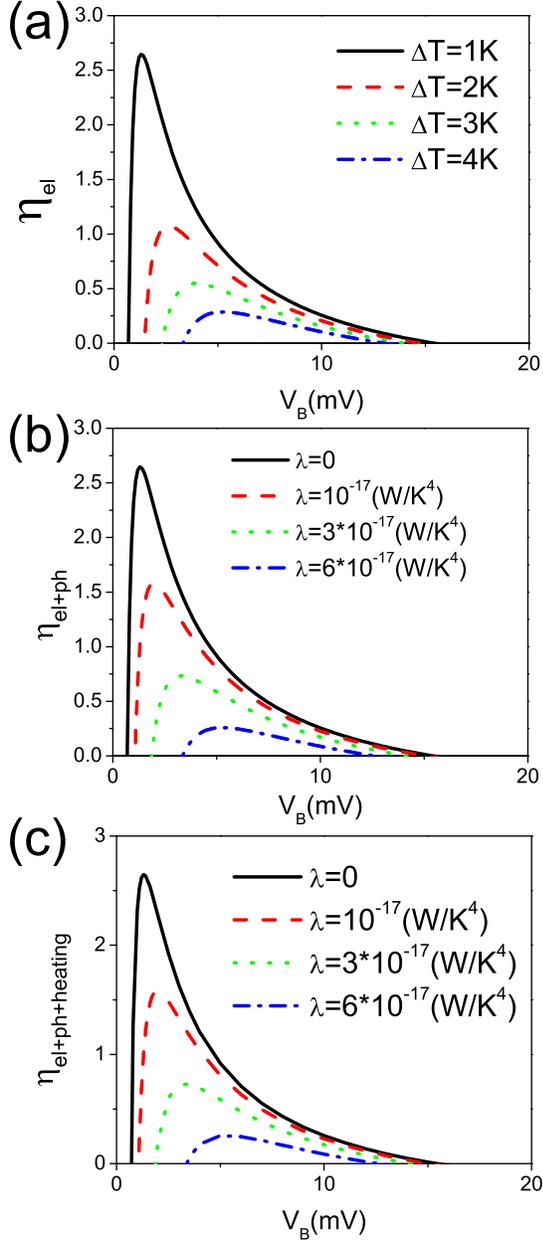}
\caption{
(color online)
(a) $\eta_{el}$ vs. $V_{B}$ in the absence of the phonon's thermal
current for , where where $T_{R}=200$~K;
(b) $\eta_{el+ph}$ vs. $V_{B}$ in the presence of the phonon's thermal
current for $\lambda=0$, $1\times10^{-17}$, $3\times10^{-17}$,
and $6\times10^{-17}$~W/K$^{4}$, where where $\Delta T=1$ K and $T_{R}=200$~K;
(c) $\eta_{el+ph+heating}$ vs. $V_{B}$ in the presence of local heating and the
phonon's thermal current for $\lambda=0$, $1\times10^{-17}$, $3\times10^{-17}$,
and $6\times10^{-17}$~W/K$^{4}$, where where $\Delta T=1$ K and $T_{R}=200$~K.
}%
\label{Fig7}%
\end{figure}
\begin{figure}[ptb]
\includegraphics[width=7.5cm]{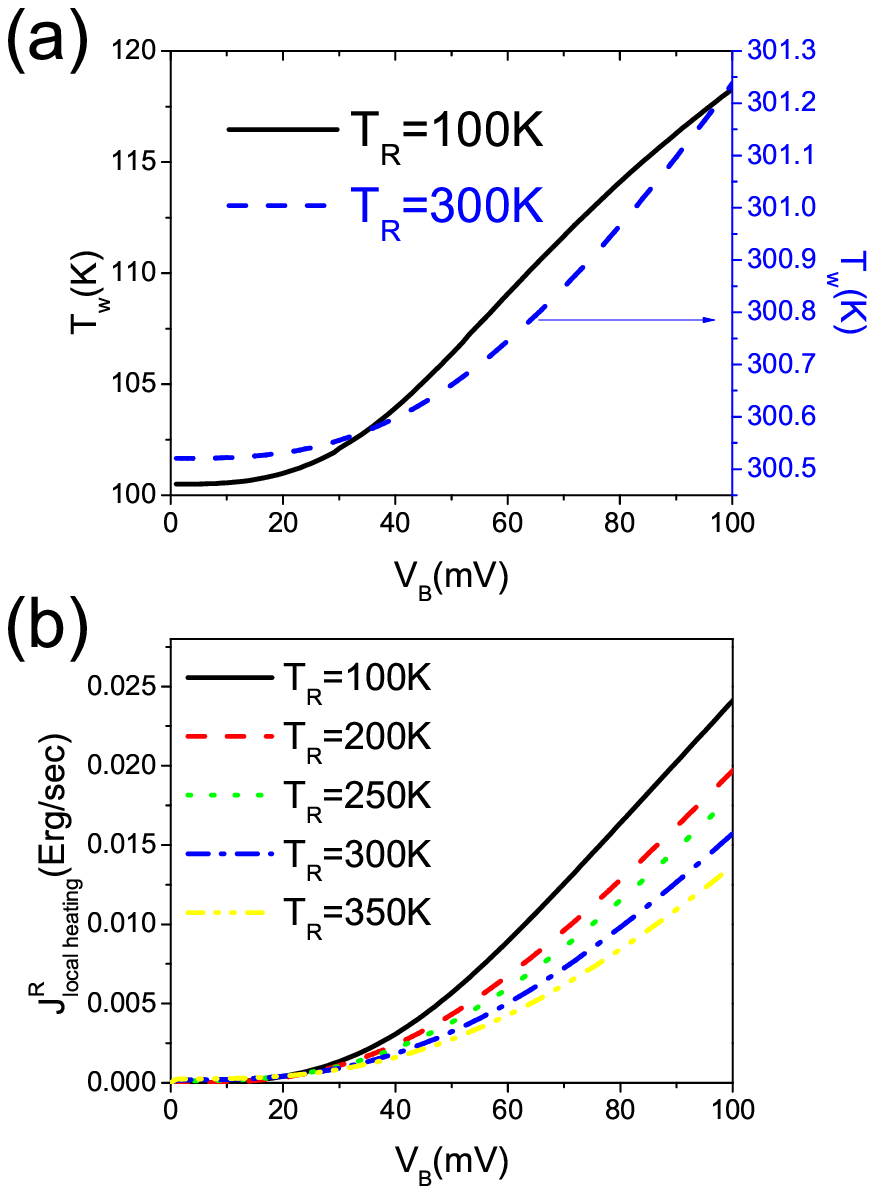}
\caption{
(color online)
(a) Effective wire temperature $T_{w}$ vs. $V_{B}$ for
$T_{R}=100$~K (left axis) and $T_{R}=300$~K (right axis), where
$\lambda=1\times10^{-17}$~W/K$^{4}$;
(b) The rate of heating energy dissipated to the cold reservoir
$J_{local~heating}^{R}$ vs. $V_{B}$ for
$T_{R}=100$, $200$, $250$, $300$, and $350$~K
}%
\label{Fig8}%
\end{figure}
This study proposes a thermoelectric cooling device based on an atomic-sized
junction. We developed an analytical theory and first-principles calculations
for the electronic cooling including the effect of the phonon's thermal
current and local heating. The theory is applied to investigate an ideal
4-Al monatomic chain sandwiched between two bulk Al electrodes. In the
Sec.~III.~A., we discuss the enhanced Seebeck effect due
to the $P_{x}-P_{y}$ orbital near the
chemical potential and the influence of the phonon's thermal current on $ZT$.
A large Seebeck coefficient is of crucial importance for the design of
thermoelectric nano-devices. Therefore, we investigate the thermoelectric
properties of the 4-Al atomic junction as follows. In the Sec.~
III.~B, III.~C, and III.~D, we discuss the thermoelectric cooling effect
without and with the phonon's thermal current and local heating. The 4-Al
thermoelectric junction serves as an example illustrating the advantage of
nano-scale refrigerators, where the overwhelmingly Joule heat in the bulk
system and photon radiation are strongly suppressed due to size reduction.
These properties facilitate possible thermoelectric cooling in the 4-Al
junction beyond the expectation of conventional solid-state device theory.
\subsection{Thermoelectric Properties of the 4-Al Atomic Junction}
We apply the theory and first-principles calculations shown in Sec.~II
to a 4-Al atomic junction as a thermoelectric cooling device, as
depicted in Fig.~\ref{Fig1}(a). The 4-Al atomic junction is electronically
simple such that the first-principle calculations reported here can be
performed with a high level of accuracy. The aluminum junction is, therefore,
an ideal testbed for comparing quantum transport theory under
non-equilibrium and experiments~\cite{Cuevas,ThygesenAl}.

We begin our discussion by considering an ideal 4-Al atomic chain bridging two
bulk Al metal electrodes that we model as ideal metals (jellium model,
$r_{s}\approx2$). The nano-structured object is considered as a scattering center.
The scattered wave functions of the whole system are calculated
by solving the Lippmann-Schwinger equation iteratively until self-consistency
is obtained. We assume that the left and right electrodes are independent
electron and phonon reservoirs ($T_{R}=T_{C}$; $T_{L}=T_{R}+\Delta T=T_{H}$),
respectively, with the electron and phonon population described by the
Fermi-Dirac and Bose-Einstein distribution function. Two electrodes are
connected to a battery with an applied bias $V_{B}=(\mu_{R}-\mu_{L})/e$, where
$\mu_{L}$ and $\mu_{R}$ are the chemical potential deep in the left and right
electrodes, respectively. A detailed account of the theory is given in Sec.~II.~A.

The Seebeck coefficients are calculated from first-principles using the
transmission function obtained from the DFT calculations, as described in
Eq.~(\ref{S}). The 4-Al atomic junction is marked by a sigma channel near the
chemical potentials with the $P_{x}-P_{y}$ orbital characters, as shown in
Fig.~\ref{Fig1}(b). The Seebeck coefficients correlate highly with the
magnitudes and slopes of DOSs near the chemical potentials, as described in
Eq.~(\ref{SlowT}). The sigma channel leads to a large value in the slope of
the transmission function near the chemical potentials. This give rise to a
larger Seebeck coefficient, as shown in Fig.~\ref{Fig1}(c), which compares
favorably with those of Pt, Pd, and Au atomic chains.

The efficiency of energy conversion in the thermoelectric junction is usually
described by the figure of merit (denoted as $ZT$), as defined in
Eq.~(\ref{zt}). When $ZT$ tends to infinity, the thermoelectric efficiency
will reach the Carnot efficiency, the upper limit of energy conversion
efficiency. $ZT$ depends on the following physical factors: the Seebeck
coefficient ($S$), the electric conductance ($\sigma$), the electron's thermal
conductance ($\kappa_{el}$), and the phonon's thermal conductance
($\kappa_{ph}$). These physical factors can be evaluated using Eqs.~(\ref{S}),
(\ref{cond}), (\ref{thercond}), and (\ref{simplifiedJph}), respectively. To
obtain a large $ZT$ value, the thermoelectric nanojunction will require a
large value of $S$, a large value of $\sigma$, and a small value of the
combined heat conductance ($\kappa=\kappa_{el}+\kappa_{ph}$). Thermoelectric
devices with a large value of $\sigma$ are usually accompanied by a large
value of $\kappa_{el}$, due to the same proportionality with the transmission
function. These values are highly correlated, making the enhancement of the
thermoelectric figure of merit $ZT$ a challenging task.

The thermal energy carried by phonons flows from the hot into cold reservoir.
The phonon's thermal current takes heat into the cold reservoir and it ,thus,
is a negative effect to the thermoelectric refrigeration. We consider the
phonon's thermal current in the weak link model [as described in
Eq.~(\ref{JQ})], where each electrode is assumed to be in thermodynamic
equilibrium, joined by a weak mechanical link modeled by a harmonic spring of
stiffness $K$. In the weak-tunneling limit, the weak link model can be
interpreted as an application of the thermal Landauer formula~\cite{Patton}.
Up to the leading order in temperatures, we expand Eq.~(\ref{JQ}) which gives
a simple form, $J_{ph}^{R}=\lambda(T_{R}^{4}-T_{L}^{4})$ [i.e.,
Eq.~(\ref{simplifiedJph}) specified as the simplified weak link
model], where $\lambda\approx2\pi^{5}K^{2}C^{2}k_{B}^{4}/(15\hbar)$.
Simplification allows us to investigate the effect of the phonon's thermal
current with a single parameter $\lambda$, which is determined by the
stiffness of the nano-structured object ($K$) and the slope of the
electrodes's surface phonon's spectral density ($C$). The simplified
weak link model is valid in small temperature regime ($T\ll T_{D}$, where
$T_{D}=394$~K is the Debye temperature for Al). We note that the
simplified weak link model violates the Wiedemann-Franz law in
low-temperature regime, while Eq.~(\ref{JQ}) restores the Wiedemann-Franz law
at $T\gg T_{D}$. The range of temperatures discussed in this study mostly lies
within the valid regime of the simplified weak link model. Therefore, we
consider $ZT$ as a function of temperatures in the presence of the phonon's
thermal current of which is represented by various values of $\lambda$, as
shown in Fig.~\ref{Fig1}(d). The validity of the $\lambda$ values for linear
atomic wires will be justified later in Sec.~III.~C.
\subsection{Thermoelectric Cooling Effect in the Absence of Phonon's Thermal
Current and Local Heating}
Let us now attempt to investigate the 4-Al junction as a thermoelectric
cooling device from the first-principles approaches. For the present, we shall
confine our attention to the simplest case, the one which neglects the
phonon's thermal current and local heating. Effects of the phonon's
thermal current and local heating will be discussed in the Sec.~III.~C
and III.~D. We also put forward an analytical theory to explain the
numerical results. The analytical theory described below is general to any
atomic/molecular thermoelectric junction as a thermoelectric nano-refrigerator.

When an external bias is applied, electrons flow from the right to the left
electrodes. The flow of electrons carries not only the charge current but also
the energy current. The thermal current carried by electrons [denoted as
$J_{el}^{R(L)}(\mu_{L},T_{L};\mu_{R},T_{R})$], defined as the rate at which
thermal energy flows from the right (into the left) electrode, can be
calculated from first-principles using Eq.~(\ref{JJ}) with the help of the
wavefunctions obtained self-consistently in the DFT calculations. Note that
$J_{el}^{R}+IV_{B}=J_{el}^{L}$ [for details see Eq.~(\ref{energyconversation}%
)] because of energy conservation. It implies that the thermoelectric
refrigeration requires electric power $IV_{B}$ to remove thermal energy from
the cold (right) reservoir (with rate $J_{el}^{R}$) and reject waste thermal
energy to the hot (left) reservoir (with rate $J_{el}^{R}$). The
nano-refrigerator works when $J_{el}^{R}>0$. As noted in Eq.~(\ref{JJ}),
$J_{el}^{R}$ is a function of $V_{B}$, $\Delta T$, and $T_{R}$. Note that
$J_{el}^{R}(T_{R}=0)<0$. At a given $V_{B}$ and $\Delta T$, the smallest
solution of $J_{el}^{R}(T_{R})=0$ defines the threshold operation temperature,
denoted as $T_{el}^{OP}$, below which the nano-refrigerator does not function.
Figure~\ref{Fig3}(a) shows the $T_{el}^{OP}$ of the 4-Al nano-refrigerator as
a function of $\Delta T$ and $V_{B}$. The operating temperature of the 4-Al
nano-refrigerator can be lower than $100$~K, as shown in Fig.~\ref{Fig3}(a).

To enrich the understanding of thermoelectric nano-refrigerators, we also
propose an analytical theory that provide guidelines for the design of
nano-refrigerators. When $\Delta T\ll T_{R}$, the application of the
Sommerfeld expansion on Eq.~(\ref{JJ}) can simplify the electron's thermal
current $J_{el}^{R}$ as a polynomial of $V_{B}$, as described in
Eq.~(\ref{JRpoly}). If the higher-order term $V_{B}^{3}$ is neglected, the
maximum of $J_{el}^{R}(T_{R})$ that is greater than zero yields a criterion
for the existence of thermoelectric cooling, which gives Eq.~(\ref{criterion}%
): $-S>\pi k_{B}\sqrt{2\Delta T/(3T_{C})}/e$, where $T_{C}$ ($=T_{R}$) is the
temperature of the cold (right) reservoir, $S=-\pi^{2}k_{B}^{2}\tau^{\prime
}(\mu)T_{R}/[3e\tau(\mu)]$ is the Seebeck coefficient, $k_{B}$ is the
Boltzmann constant, and $e$ is the electron charge. To have possible
refrigeration effect, the Seebeck coefficient $S$, the temperature difference
$\Delta T$, and the temperature of the cold reservoir $T_{C}$ need to satisfy
Eq.~(\ref{criterion}). The choice of an n-type thermoelectric junction with a
large Seebeck coefficient is of crucial importance in the design of
thermoelectric nano-refrigerators. Large Seebeck coefficients could be
achieved through an appropriate choice of bridging nano-structured objects and
further optimized by applying gate voltages~\cite{Wang,Liu}.

Figure~\ref{Fig4}(a) shows the rate of thermal energy extracted from the cold
temperature reservoir ($J_{el}^{R})$ as a function of $V_{B}$ for $T_{R}=200$,
$250$, $300$, and $350$~K, where the temperature difference between the hot
and cold reservoir $\Delta T$ is fixed at $1$~K. We note that $J_{el}^{R}<0$
at zero bias [see Fig.~\ref{Fig4}(a)], and thus a lower threshold voltage
$V_{el}^{th,lower}$ for the battery is needed to trigger the refrigeration
effect. The lower threshold bias is defined as the smallest positive solution
of $J_{el}^{R}(V_{B})=0$ for a given $T_{R}$ and $\Delta T$. When
$V_{B}<V_{el}^{th,lower}$, the rate of thermal energy removed from the cold
reservoir is negative ($J_{el}^{R}<0$) and thus the thermoelectric
nano-refrigerator does not function. We also observe an upper threshold
voltage $V_{el}^{th,upper}$ for the refrigeration effect. When $V_{B}%
>V_{el}^{th,upper}$, the rate of thermal energy removed from the cold
reservoir is also negative ($J_{el}^{R}<0$) and thus the thermoelectric
nano-refrigerator loses the capability of refrigeration. Since $V_{el}%
^{th,lower}$ and $V_{el}^{th,upper}$ have very small values, we neglect the
terms higher than the second order of $V_{B}$ in Eq.~(\ref{JRpoly}) and obtain
$V_{el}^{th,lower}$ and $V_{el}^{th,upper}$ from $J_{el}^{R}(V_{B})=0$, as
described in Eqs.~(\ref{Vth_el2}) and (\ref{Vupper_el}), where we assumed
$\Delta T\ll T_{R}$. Equation~(\ref{Vth_el2}) predicts that the lower
threshold voltage, $V_{el}^{th,lower}\approx-\pi^{2}k_{B}^{2}\Delta
T/(3e^{2}ST_{R})$, decreases as $T_{R}$ increases, as shown in the inset of
Fig.~\ref{Fig4}(a). Equation (\ref{Vupper_el}) predicts that the upper
threshold voltage, $V_{th,upper}^{el}\approx2ST_{R}^{2}$, increases as $T_{R}$
increases. Concluding from Eqs.~(\ref{Vth_el2}), (\ref{Vupper_el}) and
(\ref{eta_el_max2}), the refrigeration effect is triggered at around
$V_{B}=V_{el}^{th,lower}$, optimized around $V_{B}=2V_{el}^{th,lower}$, and
loses the refrigeration capability at $V_{B}=V_{el}^{th,upper}$.

Figure~\ref{Fig5}(a) shows $J_{el}^{R}$ as a function of $T_{R}$ for different
values of $\Delta T=0$, $1$, $2$, $3$, and $4$~K, where the bias is fixed at
$V_{B}=6$ mV. For a given $V_{B}$ and $\Delta T$, we define the critical
operation temperature $T_{el}^{OP}$ as the solution of $J_{el}^{R}(T_{R})=0$.
The thermoelectric nano-refrigerator is working when $T_{R}>T_{el}^{OP}$. The
critical operation temperature can be calculated from Eq.~(\ref{JRe1}):
$T_{el}^{OP}\approx\frac{1}{2}[(\alpha-1)\Delta T+\sqrt{(\alpha^{2}-1)\Delta
T^{2}+\beta}]$, where $\alpha=\tau(\mu_{L})/[\tau^{\prime}(\mu_{L})eV_{B}]$
and $\beta=-2V_{B}/S$. The foregoing equation predicts that $T_{el}^{OP}$
increases as $\Delta T$ increases, which agrees well with the numerical
calculation presented in Fig.~\ref{Fig5}(a). If $\Delta T=0$~K, then
$T_{el}^{OP}$ reaches the minimum value $\left(  T_{el}^{OP}\right)  _{\min
}\approx\sqrt{T_{R}V_{B}/(-2S)}$ as described in Eq.~(\ref{Top_min}). The
equation shows that $\left(  T_{el}^{OP}\right)  _{\min}$ increases as $V_{B}$
increases and as $T_{R}$ increases, respectively. At $V_{B}=6$ mV, the
$\left(  T_{el}^{OP}\right)  _{\min}$ for the 4-Al thermoelectric refrigerator
is $116$ K, which can be further suppressed by decreasing the bias $V_{B}$ .
Turning to the efficiency of the thermoelectric nano-refrigerator, the COP
(denoted as $\eta^{el}$) is defined as the ratio of the rate of heat removed
from the cold reservoir to the electric power supplied by the battery, that
is, $\eta_{el}=J_{el}^{R}/(IV_{B})$, as equivalent to Eq.~(\ref{COP2}).
Derived from Eq.~(\ref{COP4}), the maximum value of COP is given by
Eq.~(\ref{eta_el_max2}).

Figure~\ref{Fig6}(a) shows the numeric calculations of $\eta_{el}$ as a
function of $V_{B}$ for $T_{R}=200$, $250$, $300$, and $350$~K, where $\Delta
T=1$ K. This figure also shows that the optimized COP ($\eta_{el}^{\max}$)
increases as $T_{R}$ increases. The optimized COP occurs at bias $V_{\max
}^{\eta_{el}}\approx2V_{el}^{th,lower}$, where $V_{el}^{th,lower}$ is the
lower threshold bias for possible refrigeration, as shown in Fig.~\ref{Fig7}(a).
We should note that the maximum value of COP can be greatly
magnified by a suitable nanojunction with a large Seebeck coefficient according
to $\eta_{el}^{\max}\propto S^{2}$ as predicted in Eq.~(\ref{eta_el_max2}).

Figure~\ref{Fig7}(a) shows the numeric calculations of $\eta_{el}$ as a
function of $V_{B}$ for $\Delta T=1$, $2$, $3$, and $4$~K, where $T_{R}%
=200$~K. The exchange of energy by heat currents between the hot and cold
reservoirs is an irreversible process, and leads the decrease of the optimized
COP as $\Delta T$ increases as described by Eq.~(\ref{eta_el_max2}):
$\eta_{el}^{\max}=3e^{2}S^{2}T_{R}/(4\pi^{2}k_{B}^{2}\Delta T)-1/2$. As a
direct consequence of the above equation, the maximum value of COP ($\eta
_{el}^{\max}$) increases as $T_{R}$ increases and as $\Delta T$ decreases.
This prediction agrees well with Eq.~(\ref{eta_el_max2}).
\subsection{Thermoelectric Cooling Effect including the Phonon's Thermal
Current}
The phonon's thermal current carries thermal energy from the hot to cold
reservoir, which is an adverse effect to refrigeration. To assess the extent
of this adverse effect, we consider the phonon's thermal current within the
simplified weak link model. The simplified weak link model
is suitable for describing the heat transport for two thermal reservoirs
connected by a weak elastic link in a low temperature regime ($T\ll T_{D}$,
where $T_{D}=394$~K is the Debye temperature for Al) and is convenient to
describe the phonon's thermal current by a single parameter $\lambda$:
$J_{ph}^{R}=\lambda(T_{R}^{4}-T_{L}^{4})$, where $\lambda\approx2\pi^{5}%
K^{2}C^{2}k_{B}^{4}/(15\hbar)$. The parameter $\lambda$ can be determined by
$K$ and $C$, where $K$ is the stiffness of the nano-structured object
connecting to electrodes and $C$ is the slope of the spectral density
$N(E)\approx C\times E$. The simplified weak link model allows the
construction of an analytical theory that offers a concise explanation for how
the phonon's thermal current (represented by a single parameter $\lambda$)
affects thermoelectric cooling.

From recent experiments~\cite{Shiota}, it is observed that the stiffness of
the linear Pt atomic chains varies in a wide range of several orders of
magnitudes. This experiment suggests that the stiffness ($K\approx
0$ to $1.2$~N/m) is likely to depend strongly on the detailed atomic
structure of the full system, especially in the contact region. For example,
the monatomic chain could be particularly stiff along the chain direction when
the atomic chain forms a perfectly straight line; otherwise, the atomic
(zigzag) chain could easily bend with much smaller stiffness. In this view,
the magnitudes of phonon's thermal current could possibly vary in a wide range
in nanojunctions formed by monatomic chains. This feature could allow the
possibility (for example, zigzag chain instead of perfect linear chain) to
suppress the phonon's thermal current by creating a suppression in the
mechanic link. An estimation from the experimental data of $K=0$ to $1.2$~N/m
and $C\approx1.887\times10^{8}$~cm$^{2}$/erg$^{2}$ in the Pt atomic
junction yields the values of $\lambda$ range to be from $0$ to $2.05\times
10^{-19}$~W/K$^{4}$~~\cite{Shiota,Kern}.

In the 4-Al atomic junction, we choose $\lambda$ around the order of
$10^{-17}$~W/K$^{4}$ to present the strength of the phonon's thermal current.
In this range of $\lambda$, the effect of the phonon's thermal current on the
thermoelectric refrigeration is salient. For the 4-Al monatomic junction,
$\lambda\approx5.7\times10^{-15}$~Watt/K$^{4}$, which is given by
$K\approx90.91$ N/m obtained from total energy calculations and $C\approx
7.62\times10^{8}$ cm$^{2}$/erg$^{2}$ from the surface phonon dispersion
relation obtained from first-principle calculations~\cite{Chulkov}. However,
this value is notably larger than the Pt monatomic chains with $\lambda$
ranging from $0$ to $2.05\times10^{-19}$~W/K$^{4}$. The reason for this is
that the stiffness of a 4-Al linear atomic chain is $K=90.91$~N/m, which is
notably larger than the stiffness of Pt chain ($K=0$ to $1.2$~N/m)
measured in the experiments. This experiment for Pt monatomic chain infers
that the stiffness $K$ for the 4-Al junction could be overestimated by the
total energy calculations, where the chain is assume to be perfectly linear.
In this viewpoint, the $\lambda$ (around $10^{-17}$~W/K$^{4}$) we choose to
present the strength of the phonon's thermal current for the 4-Al atomic
junction could be reasonable.

In the presence of the phonon's thermal current, the COP is defined as
$\eta_{el+ph}=\frac{J_{el+ph}^{R}}{IV_{B}}$, where $J_{el+ph}^{R}=J_{el}%
^{R}+J_{ph}^{R}$ is the combined thermal current. In the presence of the
phonon's thermal current, the functioning of the 4-Al atomic refrigerator
requires higher operation temperatures (denoted as $T_{el+ph}^{OP}$), as shown
in Fig.~\ref{Fig3}(b), where we plot the operation temperature $T_{el+ph}%
^{OP}$ as a function of $V_{B}$ and $\Delta T$ with the strength of the
phonon's thermal current described by the simplified weak link model with
$\lambda=10^{-17}$~W/K$^{4}$. Applying the Sommerfeld expansion, the combined
thermal current $J_{el+ph}^{R}$ can be expressed as a polynomial of $V_{B}$
when $\Delta T\ll T_{R}$, as shown in Eq.~(\ref{JRpolyph}). When the term
$V_{B}^{3}$ is neglected in Eq.~(\ref{JRpolyph}), the criterion for the
existence of electronic cooling is given by Eq.~(\ref{criterion_ph}):
$-S>\sqrt{\left(  2\pi^{2}k_{B}^{2}/(3e^{2})+8\lambda/\sigma\right)  \left(
\Delta T/T_{R}\right)  }$. The criterion for possible thermoelectric
refrigeration provides the guideline for devising experiments to test the
effect of thermoelectric cooling at the atomic level. When $\lambda
\rightarrow0$, Eq.~(\ref{criterion_ph}) restores Eq.~(\ref{criterion}) in the
absence of the phonon's thermal current.

Figure~\ref{Fig4}(b) shows the combined thermal current $J_{el+ph}^{R}$ as a
function of $V_{B}$ for $\lambda=0$, $1\times10^{-17}$, $3\times10^{-17}$, and
$6\times10^{-17}$~W/K$^{4}$, where we fix $\Delta T=1$ K and $T_{R}=300$~K.
Note that $J_{el+ph}^{R}<0$ at zero bias. Similar to the previous discussions
in the absence of the phonon's thermal current, the thermoelectric
nano-refrigerator works when $J_{el+ph}^{R}>0$. This leads to the lower and
upper bounds of threshold voltage, as given in Eqs.~(\ref{Vth_elph2})] and
(\ref{Vupper_elph2}). The working bias which allows the operation of the
thermoelectric nano-refrigerator is restricted to a small range of bias:
$V_{el+ph}^{th,lower}<V_{B}<V_{el+ph}^{th,upper}$. Eqs.~(\ref{Vupper_elph2})
and (\ref{Vupper_elph2}) show that the lower (upper) threshold bias
$V_{el+ph}^{th,lower}$ ($V_{el+ph}^{th,upper}$) increases (decreases) by a
value of $\lambda\left[  4T_{R}^{2}\Delta T/\left(  -S\right)  \sigma\right]
$ as the intensity of the phonon's thermal current ($\lambda$) increases. This
leads to suppression of the range of working biases by the phonon's thermal
current, which agrees well with the numerical calculations as shown in
Fig.~\ref{Fig4}(b).

Figure~\ref{Fig5}(b) shows the combined thermal current $J_{el+ph}^{R}$ as a
function of $T_{R}$ for $\lambda=0$, $1\times10^{-17}$, $3\times10^{-17}$, and
$6\times10^{-17}$~W/K$^{4}$, where we fix $V_{B}=6$ mV and $\Delta T=1$ K.
Figure~\ref{Fig5}(b) exhibits the combined thermal current $J_{el+ph}^{R}$,
which could become negative at high temperatures regime. The reason for this
is that the phonon's thermal current brings heat from the hot to the cold
reservoir. Thus, the combined thermal current becomes a negative value, which
disables refrigeration capability at high temperatures. This imposes an upper
limit for the working temperature, above which $J_{el+ph}^{R}$ turns to be
negative and consequently the nano-refrigerator does not function. The
operation of nano-refrigerator is limited to a range of temperatures between
$T_{low,el+ph}^{OP}$ and $T_{high,el+ph}^{OP}$, as described in Eqs.~(\ref{Topphlow})
and (\ref{Topphhigh}). As $\lambda$ increases, Eqs.~(\ref{Topphlow})
and (\ref{Topphhigh}) predict that the range of operation
temperatures shrinks, which agrees well with the numerical calculations as
shown in Fig.~\ref{Fig5}(b).

Figure~\ref{Fig6}(b) shows the numeric calculations of $\eta_{el+ph}$ as a
function of $V_{B}$ for $T_{R}=200$, $250$, $300$, and $350$~K, where $\Delta
T=1$ K is fixed. Equation~(\ref{eta_elph_max2}) shows the upper limit of COP
(denoted as $\eta_{el+ph}^{\max}$) is given by : $\eta_{el+ph}^{\max}=\left[
\left(  \eta_{el}^{\max}\right)  ^{-1}+\lambda\left(  16T_{R}\Delta
T/(S^{2}\sigma)\right)  \right]  ^{-1}$. \ It shows that the upper limit of
COP ($\eta_{\max}^{el+ph}$) increases as $\lambda$ increases as predicted by
Eq.~(\ref{eta_elph_max2}); meanwhile, the inset of Fig.~\ref{Fig6}(b) shows
that the $V_{el+ph}^{th,lower}$ increases as $T_{R}$ increases, as predicted
by Eq.~(\ref{Vth_elph2}) valid for $\Delta T\ll T_{R}$.
Equation~(\ref{eta_elph_max2}) also predicts that and $\eta_{\max}^{el+ph}%
$\ increases as $S^{2}\sigma$ decreases, indicating that a large value of
Seebeck coefficient is of critical importance for efficient thermoelectric refrigeration.

Figure~\ref{Fig7}(b) shows the numeric calculations of $\eta_{el+ph}$ as a
function of $V_{B}$ for $\lambda=0$, $1\times10^{-17}$, $3\times10^{-17}$, and
$6\times10^{-17}$~W/K$^{4}$, where $\Delta T=1$ K and $T_{R}=200$~K.It shows
that the upper limit of COP ($\eta_{el+ph}^{\max}$ ) decreases as the strength
of phonon's thermal current increases because the phonon's thermal current is
an adverse effect to refrigeration. Figure~\ref{Fig7}(b) also shows that the
lower threshold bias $V_{el+ph}^{th,lower}$ increases as $\lambda$ increases
and agrees well with the prediction of Eq.~(\ref{Vth_elph2}).

In short, the phonon's thermal current, which is relevant to the mechanical
coupling between the nano-structured object and the electrodes, is an adverse
effect to refrigeration. To minimize the adverse effect, we suggest creating
a weak mechanical link between the nano-structured object and the electrodes
while still allowing electrons to tunnel.

\subsection{Thermoelectric Cooling Effect Including the Phonon's Thermal
Current and Local Heating}

We further consider the effect of local heating on the thermoelectric refrigeration.
Electrons that travel with energies
larger than the energy of normal modes can excite corresponding vibrations in
the nano-structure anchoring the electrodes. This effect causes local heating
in the nano-structure~\cite{chenheating,Frederiksen,Huang2}. Local heating occurs
when electrons exchange energy with the excitation and relaxation of the
energy levels of the vibration of the nano-structured object that anchors the
electrodes. The nano-object bridging the junction is formed by few atoms, and
thus the dispersion relation of phonon is characterized by the lack of
Goldstone mode. When normal coordinates are considered, the complex vibrations
of nano-object connecting to heavy electrodes can be cast into a set of
independent simple harmonic oscillators described as normal modes. Due to the
selection rule, the contributions to local heating from modes with large
vibrational components along the direction of propagating electrons are
important~\cite{chenh2}. The smallest longitudinal normal mode has an energy
of $eV_{onset}\approx20$ meV~\cite{Yang}.

The heat generated in the central wire region can be dissipated to the bulk
electrodes via phonon-phonon interactions. The heat generation eventually
equilibrates the heat dissipation, where the wire region reaches an effective
local temperature $T_{w}$ higher than the averaged electrode temperatures
$(T_{L}+T_{R})/2$. Local temperature depends on several factors: the strength
of coupling between electrons and the vibrations, the background temperature,
and the mechanical coupling between the nano-structure and electrodes which
determines the efficiency of thermal current dissipating the thermal energy of
local heating.

In Fig.~\ref{Fig8}(a) we show the effective wire temperature ($T_{w}$) as a function
of bias ($V_{B}$) for $T_{R}=100$ and $300$~K, where we assume that the
strength of the phonon's thermal current is given by $\lambda=1\times10^{-17}%
~$W/K$^{4}$. For $T_{R}=100$ K, the increase of local wire temperature $T_{w}$
is noticeable when $V_{B}>V_{onset}\approx20$~mV. We note that $T_{w}$ is
slightly higher than $T_{R}$ when $V_{B}<V_{onset}$ because only a small
portion of electrons has energy larger than $eV_{onset}$ due to the tail of
the Fermi-Dirac distributions at finite temperatures. For $T_{R}=300$~K, the
increase in local wire temperature $T_{w}$ is significantly suppressed due to
increasingly efficient heat dissipation at high temperatures. The local
heating generated in the wire region is dissipated to the left (hot) and
right (cold) electrodes as described in Eqs.~(\ref{JLheating}) and
(\ref{JRheating}), respectively. The combined thermal current including
the effect of the phonon's thermal current and local heating ($J_{el+ph+heating}%
^{R}$) is given by Eq.~(\ref{JR_elphheating}).

In Fig.~\ref{Fig8}(b)we show the rate of local heating energy which is dissipated
to the cold (right) electrode ($J_{local~heating}^{R}$) as a function
of bias ($V_{B}$) for $T_{R}=100$, $200$, $250$, $300$, and $350$~K, where we
assume that the strength of the phonon's thermal current is given by
$\lambda=1\times10^{-17}$~W/K$^{4}$. The rate of heating energy dissipated to
the right (cold) electrode ($J_{local~heating}^{R}$) introduces an
additional negative effect to refrigeration, and, thus, the combined thermal
current $J_{el+ph+heating}^{R}$ which takes heat energy from the right (cold)
reservoir to left (hot) reservoir is suppressed. In this bias range, the
heating power is smaller than 1\% of the electric power ($IV_{B}$) supplied by
a battery even at ambient temperatures because of the quasi-ballistic
transport. This fact leads to  $J_{local\text{ }heating}^{R}\ll J_{ph}^{R}$ in
the range of working biases ($V_{el+ph}^{th,lower}<V_{B}<V_{el+ph}^{th,upper}%
$) for the $\lambda$ values we considered in this study. Consequently,
the effect of local heating on thermoelectric refrigeration is negligible when the
phonon's thermal current is included. This finding has profound implications
on the design of nanoscale thermoelectric refrigerator, where the local
heating is strongly suppressed such that the overwhelming Joule heating in the
bulk system can be avoided. This feature remarkably facilitates the electron
cooling beyond the expectation of the conventional thermoelectric device
theory.

We repeat the calculations in the previous two subsections. The results are not
visibly different when we compare the ones that include contributions from both
phonon's thermal current and local heating, as shown in Figs.~\ref{Fig4}(c),
\ref{Fig5}(c), \ref{Fig6}(c), and \ref{Fig7}(c), with the ones that has phonon's
thermal current only, as shown in Figs.~\ref{Fig4}(b), \ref{Fig5}(b),
\ref{Fig6}(b), and \ref{Fig7}(b).
\section{Conclusions}
In summary, we propose a thermoelectric nano-refrigerator based on an atomic
junction, the extreme limit of device minimization. In-depth research which
combines the first-principles and analytical calculations is developed to
study the thermoelectric cooling mechanism. The theory is applied to
investigate the thermoelectric cooling of the 4-Al monatomic junction. The
4-Al junction is electronically simple such that the first-principle
calculations reported here can be performed with a high level of accuracy. It,
thus, serves an ideal testbed for comparing the prediction of the theory
and experiments. Our studies show that the $P_{x}-P_{y}$ orbital characters near the
chemical potential lead to a larger Seebeck coefficient in the 4-Al atomic
junction. We investigated the working conditions, operation temperatures,
electron's thermal current which removes heat from the cold temperature
reservoir, and COP in the presence of the phonon's thermal current and local heating.

Firstly, we perform first-principles calculations for the electron's thermal
current ($J_{el}^{R}$) which removes heat from the cold temperature reservoir
in the absence of the phonon's thermal current and local heating. The
thermoelectric cooling is working when $J_{el}^{R}>0$. It is observed that the
operation of nano-refrigerators requires a minimum critical working
temperature. The solution of $J_{el}^{R}(T_{R})=0$ defines the critical
operation temperature denoted as $T_{el}^{OP}$. When $T_{R}<T_{el}^{OP}$,
nano-refrigerators do not function. When $\Delta T=0$, $T_{el}^{OP}$
reaches the minimum value $(T_{el}^{OP})_{\min}$, as given by Eq.~(\ref{Top_min}).
It is observed that the lowest critical operation temperature
may be smaller than $100$~K, which is very efficient as a low-temperature
operated nano-refrigerator compared with the vacuum diode due to reduced work
function via resonant tunneling. The working condition of the thermoelectric
nano-refrigerator is quite demanding. The applied voltage is restricted to a
small range of biases. It is observed that there are lower and an upper bounds
of biases denoted as $V_{th}^{el}$ and $V_{upperth}^{el}$, respectively. The
thermoelectric nano-refrigerator does not work when the applied biases are
smaller than $V_{th}^{el}$ or larger than $V_{upperth}^{el}$. Since the lower
and upper threshold biases are small, it allows us to expand $J_{el}^{R}$ and
obtain an analytical expression for $V_{th}^{el}$ and $V_{upperth}^{el}$, as
described in Eqs.~(\ref{Vth_el2}) and (\ref{Vupper_el}). The maximum value of
COP ($\eta^{el}$) occurs at bias around $2V_{th}^{el}$ and can be analytically
expressed as Eq.~(\ref{eta_el_max2}).

Secondly, we consider the phonon's thermal current, which is a major effect
against the operation of atomic refrigerators. To have a more perspective
realization, we consider the phonon's thermal current in the approximation of
the simplified weak-link model: $J_{ph}^{R}=\lambda(T_{R}^{4}-T_{L}^{4})$. The
simple model has a single parameter ($\lambda$) which allows us to construct
an analytical theory to show the effect of the phonon's thermal current on
electronic cooling in a transparent way. We choose $\lambda$ around the order
of $10^{-17}$~W/K$^{4}$ to present the effect of the phonon's thermal current.
In this range of $\lambda$, the effect of the phonon's thermal current on the
thermoelectric refrigeration is salient and sensitive. In the presence of the
phonon's thermal current, the combined thermal current ($J_{el+ph}^{R}$) which
removes heat from the cold temperature reservoir decreases as $\lambda$
increases. This leads to higher critical operation temperature $T_{el+ph}%
^{OP}$. The working conditions of the thermoelectric nano-refrigerator are
hampered by the phonon's thermal current further, as described in Eq.~(\ref{criterion_ph}).
The range of biases which are allowed to drive
nano-refrigerators shrinks. The lower bound of the operating bias
($V_{th}^{el+ph}$) increases as the intensity of the phonon's thermal current
($\lambda$) increases, as described in Eq.~(\ref{Vth_elph2}). The upper bound
of the operating bias ($V_{upperth}^{el+ph}$) decreases as $\lambda$
increases. The COP ($\eta^{el+ph}$) is also suppressed by the intensity of the
phonon's thermal current ($\lambda$) because the phonon's thermal current,
taking heat from the hot to cold temperature reservoir, is an adverse effect
to thermoelectric refrigeration. The suppression of the optimized COP
($\eta_{\max}^{el+ph}$) by $\lambda$\ agrees well with the analytical
expression given by Eq.~(\ref{eta_elph_max2}). We would like to stress that
the $\lambda$ values chosen to present the strength of the phonon's thermal
current is quite large. For example, the rate of thermal energy flows from
electrode at $100$ K to electrode at $0$ K is $1$~nW when $\lambda
=1\times10^{-17}$~W/K$^{4}$.

Thirdly, we consider the effect of local heating on the thermoelectric
refrigeration. Electrons that propagate with the energies larger than the
energies of normal modes can excite corresponding vibrations in the
nano-object anchoring the electrodes. This effect causes local heating in the
nano-structure, which is analogous to Joule heating in the bulk system caused
by diffusive electrons. The heat generated in the center region of the
junction is dissipated to the hot and cold temperature reservoirs; and, thus,
local heating is an adverse effect to electronic cooling. In the bulk system,
irreversible Joule heating is overwhelming such that the efficiency of a
thermoelectric refrigerator is significantly suppressed. Fortunately, the
quasi-ballistic nature of electron transport in the atomic scale junctions
significantly reduces local heating due the size reduction. This quantum
feature remarkably facilitate thermoelectric cooling beyond the expectation of
the conventional thermoelectric device theory. To demonstrate this point, we
perform first-principles calculations for local heating using the Fermi
golden rule in the first-order perturbation theory in the frame work of
density functional theory. Our calculations show that local heating dissipated
to the cold temperature reservoir is very small compared with the large
phonon's thermal current considered in this study. Consequently, local heating
is negligible, especially in the small bias regime where the nano-refrigerants
is functioning. Moreover, the photon radiation is also negligible even when we
consider the perfect black body radiation for a nanojunction with a surface
area of $1$ $\mu$m diameter compared with the large phonon's thermal current.

Finally, we would like to mention that the simplified weak link model may not be
perfect in quantitative description of the phonon's thermal current. For
example, it is valid only when for $T\ll T_{D}$, where $T_{D}=394$~K is the
Debye temperature for Al. The value of $\lambda\approx2\pi^{5}K^{2}C^{2}%
k_{B}^{4}/(15\hbar)$ is unknown due to the uncertainty of the stiffness $K$ of
the bridging nano-object. For the 4-Al monatomic junction, $\lambda
\approx5.7\times10^{-15}$ Watt/K$^{4}$where $K\approx90.91$ N/m is obtained
from total energy calculations and $C\approx7.62\times10^{8}$ cm$^{2}%
$/erg$^{2}$ is obtained from the surface phonon dispersion relation from
electronic-structure calculations. This $\lambda$ value is considerably larger
than the Pt monatomic chains with $\lambda$ ranging from $0$ to $2.05\times
10^{-19}$~W/K$^{4}$, where $K=0\symbol{126}1.2$ N/m is obtained from the
experiments and $C=1.887\times10^{8}$ cm$^{2}$/erg$^{2}$ is obtained from the
surface phonon dispersion relation of Pt. The reason for large discrepancy
between the $\lambda$ value is as the following: the stiffness of a 4-Al linear
atomic chain ($K=90.91$ N/m) from total energy calculations is much larger
than the stiffness of Pt monatomic chain ($K=0\symbol{126}1.2$ N/m) measured
in experiments. This comparison infers that the stiffness $K$ is likely to
depend strongly on the detailed atomic structure of the full system,
especially in the contact region. For example, the atom chain could be
particularly stiff along the chain direction when the atomic chain forms a
perfectly straight line; otherwise, the atomic (zigzag) chain could easily
bend with much smaller stiffness. The total energy calculations have assumed
that the atomic chain is perfectly linear. This may lead to significant
overestimation for the $\lambda$ value. The imperfection of the contact region
in the real system may allow the possibility (for example, zigzag chain instead
of perfect linear chain) to suppress the phonon's thermal current by creating
a frustration in the mechanical link connecting to the electrodes. In this
case, the thermoelectric refrigeration at atomic scale may be salient.

In short, atomic-level control of the contact region is expected to open new
opportunities and challenges in developing new forms of thermoelectric energy
conversion devices. Atomic scale thermoelectric devices need to be extended by
the utilization of unprecedented experiments. The nano-refrigerators
potentially have better performance than the conventional TE refrigerators
with the same $ZT$. The is due to the suppression of local heating and photon
radiation due to the small size, which avoid the overwhelming Joule heating in
the bulk system. This feature remarkably facilitates the electron cooling
beyond the expectation of the conventional thermoelectric device theory.

The authors thank MOE ATU, NCHC, National Center for Theoretical
Sciences(South), and NSC (Taiwan) for support under Grants NSC
97-2112-M-009-011-MY3, 098-2811-M-009-021, and 97-2120-M-009-005.

\end{document}